\documentclass[preprint]{ptephy_v1}

\preprintnumber{YITP-23-12; J-PARC-TH-0283} 
\usepackage{hyperref}

\usepackage{amssymb}
\usepackage{amsmath}
\usepackage{bm}
\usepackage[subrefformat=parens]{subcaption}
\usepackage{xcolor}
\usepackage{slashed}
\usepackage{hyperref}
\usepackage{ulem}
     
\hypersetup{
     colorlinks=true,       
     linkcolor=blue,          
     citecolor=blue,        
     urlcolor=blue,           
}

\begin{document}

\title{
Enhancement of dilepton production rate and electric conductivity
around QCD critical point
  }


\author[1,2]{Toru Nishimura}
\author[2,3,1]{Masakiyo Kitazawa}
\author[2]{Teiji Kunihiro}

\affil[1]{Department of Physics, Osaka University, Toyonaka, Osaka, 560-0043 Japan
\email{nishimura@kern.phys.sci.osaka-u.ac.jp}}
\affil[2]{Yukawa Institute for Theoretical Physics, Kyoto University, Kyoto, 606-8502 Japan}
\affil[3]{J-PARC Branch, KEK Theory Center, Institute of Particle and Nuclear Studies, KEK, 319-1106 Japan}


\begin{abstract}
We investigate whether the soft mode that becomes massless
at the QCD critical point (CP) causes an enhancement 
of the dilepton production rate (DPR) 
and the electric conductivity around the CP through 
the modification of the photon self-energy. 
The modification is described by the so-called
Aslamazov-Larkin, Maki-Thompson and density of states terms, 
which have been taken into account in our previous study 
on the DPR near the color-superconducting phase transition,
with a replacement of the diquark modes with the soft mode of the QCD CP.
We show that the coupling of photons with the soft modes brings 
about an enhancement of the DPR in the low invariant-mass region 
and the conductivity near the CP, 
which would be observable in the relativistic heavy-ion collisions.
\end{abstract}


\maketitle

\section{Introduction}
Exploring the high-density matter at vanishing and finite temperature 
in Quantum Chromodynamics (QCD) is one of the most challenging 
as well as intriguing subjects
in the current nuclear physics~\cite{Lovato:2022vgq}.
Among various interesting subjects, the possible existence 
of a critical point called the QCD CP on 
the QCD phase diagram has been acquiring much attention.
The phase transition at the QCD CP is of second order
with the same universality class as the $Z_2$ Ising model,
and large fluctuations of various quantities coupled to the order parameter 
are expected to occur~\cite{Stephanov:1998dy, Stephanov:1999zu}.
A number of proposals have been made for observational identification
of the QCD CP in the relativistic heavy-ion collision (HIC) experiments
~\cite{Stephanov:1998dy, Stephanov:1999zu, Hatta:2003wn,
Minami:2009hn, Asakawa:2009aj, Asakawa:2015ybt, Hasanujjaman:2020zex, Lovato:2022vgq},
such as the event-by-event fluctuations of conserved charges 
and especially their non-Gaussianity, 
large fluctuations of the low-momentum particle distributions,
anomalous fluid dynamical phenomena 
with diverging transport coefficients and so on.
Active experimental analyses are ongoing at the beam-energy scan program 
at RHIC, NA61/SHINE, and HADES~\cite{Galatyuk:2019lcf}.
The future experiments at FAIR and J-PARC-HI will further pursue them~\cite{Agarwal:2022ydl,Ozawa:2022sam}.

In this article, we investigate possible signals 
of the QCD CP that would be observed in these experiments 
on the basis of the fact that the second-order nature of the QCD CP 
implies the existence of a low-energy mode with a vanishing mass at the CP.
Such a slow mode is called the soft mode of the phase transition.
The soft mode of the QCD CP is fluctuations in the scalar channel 
but {\it not} a sigma {\it mesonic} mode.
Instead, it is the particle-hole (p-h) collective excitation
with a mixing of baryon number density and energy density
that has a spectral support in the space-like 
region~\cite{Fujii:2004jt,Son:2004iv,Yokota:2016tip,Yokota:2017uzu}.

The existence of the soft mode should affect various observables near the CP.
In this article, as examples of such observables, 
we explore how the dilepton production rate (DPR)
and the electric conductivity are affected by the soft mode of the QCD CP.
We have shown in a previous work Ref.~\cite{Nishimura:2022mku}
that the DPR can be greatly 
enhanced in the low invariant-mass region near the phase boundary 
of the two-flavor color superconductivity (2SC)
due to the diquark soft mode~\cite{Voskresensky:2003wd, Kitazawa:2001ft, 
Kitazawa:2003cs, Kitazawa:2005vr}; in Ref.~\cite{Nishimura:2022mku}, 
the enhancement of the DPR originates from a modification of the 
photon self-energy by the Aslamazov-Larkin (AL)~\cite{AL:1968}, 
Maki-Thompson~\cite{Maki:1968, Thompson:1968} 
and density of states (DOS) terms~\cite{book_Larkin} 
incorporating the diquark soft modes. 
A surprise in Ref.~\cite{Nishimura:2022mku}
was that although the spectral support of the diquark soft mode is concentrated
in the {\it space-like} region, their scattering process 
described by the AL term does cause the enhancement 
of the DPR in the {\it time-like} region.
We thus expect that such an enhancement of these observables 
may occur by a similar mechanism 
due to the soft mode associated with the QCD CP;
we consider the AL, MT and DOS terms with the diquark soft modes 
being replaced by the soft mode of the QCD CP 
in the 2-flavor Nambu--Jona-Lasinio (NJL) model.

A notable feature of the soft mode of the QCD CP 
is that its propagator is not analytic at the origin
unlike the diquark modes investigated in Ref.~\cite{Nishimura:2022mku}.
As a result, a simple time-dependent Ginzburg-Landau (TDGL)
approximation is not applicable to describe the soft mode of the QCD CP.
We thus introduce an approximation scheme that 
simply takes care of the specific analytic properties.
The vertex functions in the AL, MT and DOS terms are then constructed
so as to be consistent with this treatment in light of the gauge invariance.
In this way, our photon self-energy is constructed 
to satisfy the Ward-Takahashi (WT) identity.

Using the photon self-energy thus constructed, 
we calculate the DPR and the electric conductivity near the QCD CP.
We show that the DPR at low invariant-mass region, 
as well as the electric conductivity, 
is greatly enhanced around the QCD CP due to the soft modes.
We also present some issues which are relevant when 
pursuing an experimental measurement of these signals in the HIC experiments.

This paper is organized as follows.
In the next section, after introducing the model and
its phase diagram in the mean-field approximation,
we discuss properties of the soft mode of the QCD CP.
In Sec.~\ref{Photon self-energy}, we calculate the photon self-energy
described by the AL, MT and DOS terms.
In Sec.~\ref{Numerical results}, we discuss the numerical results 
on the DPR and the electric conductivity near the QCD CP.
The final section will be devoted to a short summary.

\section{Phase diagram and soft modes of QCD CP}
\label{Model and soft modes of QCD CP}

To investigate the DPR and electric conductivity near the QCD CP,
we adopt the following 2-flavor NJL model~\cite{Hatsuda:1994pi} 
\begin{align}
  \mathcal{L} = \bar{\psi} i  ( \slashed \partial - m ) \psi
  + \ G_S [(\bar{\psi} \psi)^2 + (\bar{\psi} i \gamma_5 \vec\tau \psi)^2],
  \label{eq:lagrangian}
\end{align}
where $\psi$ is the quark field and $\vec\tau=(\tau_1,\tau_2, \tau_3 )$ 
is the Pauli matrices for the flavor $SU(2)_f$.
The current quark mass $m=5.5~{\rm MeV}$, 
the scalar coupling constant $G_S=5.50~\rm{GeV^{-2}}$ and 
the three-momentum cutoff $\Lambda=631~{\rm MeV}$ are determined
so as to reproduce the pion mass $m_{\pi}=138~\rm{MeV}$ and
the pion decay constant $f_{\pi}=93~\rm{MeV}$~\cite{Hatsuda:1994pi}.

\begin{figure}[t]
\centering
\includegraphics[keepaspectratio, scale=0.35]{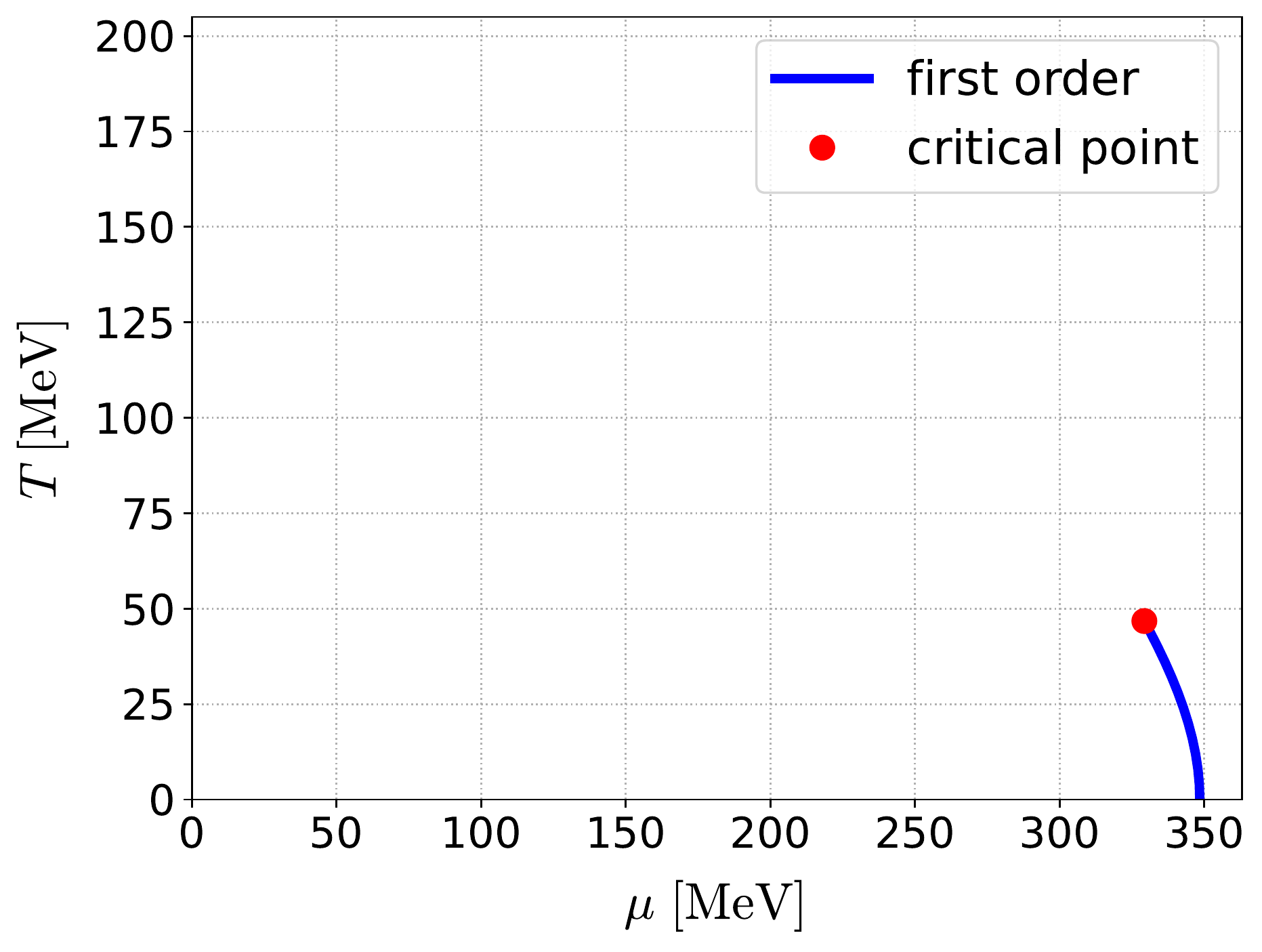}
\caption{
Phase diagram calculated by the mean-field approximation 
in the 2-flavor NJL model (\ref{eq:lagrangian}).
The solid line shows the first-order phase transition. 
The QCD CP is represented by the circle marker,
which is located at $(T_c, \mu_c) \simeq (46.757, 329.30)$~MeV.
}
\label{fig:phase-D}
\end{figure}

In Fig.~\ref{fig:phase-D}, we show the phase diagram 
as a function of the temperature $T$ and the quark chemical potential $\mu$
in the mean-field approximation 
with the mean field $\langle \bar{\psi} \psi \rangle$.
The solid line shows the first-order critical line,
and the circle marker denotes the QCD CP, which is located at
$(T_c,\, \mu_c)\simeq(46.757,\,329.30)~{\rm MeV}$. 

The soft mode of the QCD CP is described by the collective excitations 
of the scalar field $\bar\psi\psi$~\cite{Fujii:2004jt,Fujii:2003bz}.
The imaginary-time Green's function of this channel
in the random-phase approximation (RPA) is given by~\cite{Hatsuda:1994pi}
\begin{align}
  \tilde{\Xi}(k) &= \frac{1}{G_S^{-1}+\mathcal{Q}(k)},
  \label{eq:Xi} \\
  \mathcal{Q} (k) &= 2N_fN_c
  \int_p {\rm Tr} [\mathcal{G}_0 (p-k) \mathcal{G}_0(p)], 
  \label{eq:Q} 
\end{align}
where $N_f=2$ and $N_c=3$ are the numbers of flavor and color,
$\mathcal{Q}(k)=\mathcal{Q}(\bm{k}, i\nu_n)$ is 
the one-loop quark--anti-quark correlation function,
$\mathcal{G}_0(p)=\mathcal{G}_0(\bm{p}, i\omega_m) 
= 1/[(i\omega_m + \mu)\gamma_0 - \bm{p} \cdot \bm{\gamma} + M]$
 is the free-quark propagator, 
$\omega_m$ ($\nu_n$) is the Matsubara frequency for fermions (bosons), 
$M = m - 2G_S\langle \bar{\psi} \psi \rangle$ is the constituent quark mass 
and ${\rm Tr}$ is the trace over the Dirac indices.
Throughout this paper, we denote the momentum integration 
and Matsubara-frequency summation as 
$\int_p = T\sum_m \int d^3\bm{p}/(2\pi)^3$.
The Green's function $\tilde{\Xi}(k)$ is represented 
by a sum of repeated bubble diagrams composed of 
the one-loop correlation function~(\ref{eq:Q}).

The retarded functions $\Xi^R(\bm{k}, \omega)$ and $Q^R(\bm{k}, \omega)$ 
corresponding to $\tilde{\Xi}(k)$ and $\mathcal{Q}(k)$ are obtained 
by an analytic continuation $i\nu_n \rightarrow \omega+i\eta$.
The analytic formula of the imaginary part of $Q^R(\bm{k}, \omega)$ 
is calculated to be
\begin{align}
{\rm Im} Q^R (\bm{k}, \omega) 
=& -\frac{N_f N_c T}{4 \pi} \frac{\omega^2-\bm{k}^2-4M^2}{|\bm{k}|}
\nonumber \\
&\times
\Biggl\{
\theta \bigl( |\omega| - \sqrt{\bm{k}^2+4M^2} \bigr) 
F \bigl( \omega, \bar{k}(|\bm{k}|, \omega) \bigr)
\nonumber \\ 
&\quad +
\theta \bigl( \bar{k}(|\bm{k}|, \bar{\Lambda}) - |\omega| \bigr) 
\Bigl[ 
F \bigl( \omega, \bar{k}(|\bm{k}|, \omega) \bigr) - 
F \bigl( \omega, \bar{\Lambda} \bigr)
\Bigr]
\Biggr\},
\label{eq:ImQ} \\
F(\omega, x) 
=&\sum_{s, t=\pm} s~{\rm log}~{\rm cosh} \frac{\omega+sx-2t\mu}{4T},
\\
\bar{k} (|\bm{k}|, \omega)
=& |\bm{k}| \sqrt{ 1 - 4M^2/ (\omega^2-\bm{k}^2)},
\quad
\bar{\Lambda}=2\sqrt{\Lambda^2+M^2},
\end{align}
where $\bar{k}(|\bm{k}|, \bar{\Lambda}) < |\bm{k}|$.
Then the real part is given by the Kramers-Kronig relation 
\begin{align}
{\rm Re} Q^R(\bm{k}, \omega) =  \frac{1}{\pi} P 
\int^{\bar{\Lambda}}_{-\bar{\Lambda}} 
d\omega' \frac{{\rm Im} Q^R (\bm{k}, \omega')}{\omega'-\omega},
\label{eq:ReQ}
\end{align}
where $P$ denotes the principal value.

The first and second terms in the curly bracket 
in Eq.~(\ref{eq:ImQ}) take nonzero values 
in the time- and space-like regions, respectively. 
$\Xi^R(\bm{k},\omega)$ has poles that physically represent collective modes in the time- and space-like regions, respectively.
The former corresponds to the sigma meson composed of 
quark--anti-quark excitations, while the latter 
to that composed of p-h excitations due to the existence of a Fermi sphere.

From Eq.~(\ref{eq:ImQ}) one also finds that $Q^R(\bm{k}, \omega)$ 
is not analytic at the origin $(|\bm{k}|,\omega)=(0,0)$. 
In fact, the limiting value of ${\rm Im} Q^R(\bm{k}, \omega)$
at the origin along the line $\omega=a|\bm{k}|$ is given by
\begin{align}
\lim_{|\bm{k}| \to 0}{\rm Im} Q^R(\bm{k}, a |\bm{k}|) 
= a \frac{N_f N_c M^2}{2\pi} \sum_{t=\pm} 
\biggl\{
{\rm tanh} \frac{\lambda_0 -2t\mu}{4T}-{\rm tanh} \frac{\bar{\Lambda} -2t\mu}{4T}
\biggr\}
\theta\biggl(\frac{2\Lambda}{\bar{\Lambda}} - |a|\biggl) ,
\label{eq:Qlimit}
\end{align} 
with $\lambda_0 = \sqrt{4M^2/(1-a^2)}$.
Equation~(\ref{eq:Qlimit}) is nonzero for $0 < |a| < 2\Lambda/\bar{\Lambda}<1$, in which the value depends on $a$.

At the QCD CP, $\Xi^R(\bm{k}, \omega)$ satisfies
\begin{align}
{\Xi^R}^{-1} (\bm{0}, 0)\big|_{T=T_c,~\mu=\mu_c}=0,
\label{eq:Thouless}
\end{align}
in accordance with the nature of the second-order phase transition at the CP.
In fact, Eq.~(\ref{eq:Thouless}), known as the Thouless criterion~\cite{Thouless},
is derived from the stationary condition of the effective potential at the CP.
The Thouless criterion shows the existence of a collective mode 
that becomes exactly massless in $\Xi^R(\bm{k}, \omega)$.
This mode is called the soft mode associated with the CP.
It is known that the soft mode of the QCD CP is a p-h mode 
in the space-like region, while the mesonic mode in the time-like region 
does not become massless even 
at the QCD CP~\cite{Fujii:2004jt, Fujii:2003bz,Yokota:2016tip,Yokota:2017uzu}.

In the next section, we investigate the effect of the soft mode on 
the photon self-energy in the low energy and momentum region.
For this analysis we introduce an approximate formula of 
$\Xi^R (\bm{k}, \omega)$ that is valid near the QCD CP in the following way:
First, since the spectral function of the soft mode 
has the support in the space-like region, 
we focus on the strength in the space-like region only.
The mesonic mode in the time-like region is neglected since its contribution 
to the photon self-energy at low energy-momentum is suppressed 
because of the dispersion relation $\omega > \sqrt{\bm{k}^2+4M^2}$, 
where $M\simeq185~{\rm MeV}$ around the CP.
Second, we approximate the denominator of $\Xi^R (\bm{k}, \omega)$
in the space-like region by expanding it 
with respect to $\omega$ and picking up the first two terms as
\begin{align}
  \Xi^R(\bm{k}, \omega) = \frac{1}{G_S^{-1}+Q^R(\bm{k},\omega)} \sim \frac{1}{A(\bm{k}) + C(\bm{k}) \omega},
  \label{eq:XiAB}
\end{align}
where $A(\bm{k}) = G_S^{~-1} + Q^R (\bm{k}, 0)$ 
and $C(\bm{k}) = \partial Q^R (\bm{k}, \omega)/\partial \omega~|_{\omega=0}$,
which are found to be real and pure-imaginary numbers, respectively,
from Eqs.~(\ref{eq:ImQ}) and~(\ref{eq:ReQ}).
We then write the imaginary part of Eq.~(\ref{eq:XiAB}) as 
\begin{align}
  {\rm Im} \Xi^R(\bm{k}, \omega)
  \sim {\rm Im} \frac{1}{A(\bm{k}) + C(\bm{k}) \omega}
  ~\theta \bigl( \bar{k}(|\bm{k}|, \bar{\Lambda}) - |\omega| \bigr), 
\label{eq:TDGL-like}
\end{align}
where we have used the fact that ${\rm Im} \Xi^R(\bm{k}, \omega)$ takes
a nonzero value for $|\omega|<\bar{k}(|\bm{k}|, \bar{\Lambda})$ 
in the space-like region, as seen from Eq.~(\ref{eq:ImQ}).
In the next section, we use the forms of $A(\bm{k})$ and $C(\bm{k})$ 
determined in the NJL model for a given $T$ and $\mu$.
It is shown that $C(\bm{k})$ behaves as $1/|\bm{k}|$ and 
diverges in the limit $|\bm{k}|\to0$ corresponding to 
the non-analytic nature of $\Xi^R(\bm{k}, \omega)$ at the origin.
From this behavior a simple 
time-dependent Ginzburg-Landau (TDGL) approximation~\cite{book_Larkin} 
that expands $[\Xi^R(\bm{k}, \omega)]^{-1}$ 
with respect to $\omega$ and $|\bm{k}|$ is not applicable in the present case.
Our approximation~(\ref{eq:TDGL-like}) is valid 
even in this case since the $\bm{k}$ dependence is treated exactly.
We also note that the denominator of Eq.~(\ref{eq:TDGL-like})
does not diverge at the origin 
since the term $C(\bm{k})\omega$ is suppressed 
by the condition $|\omega|<|\bm{k}|$ for $\bm{k}\rightarrow0$.

\begin{figure*}[t]
    \begin{tabular}{cc}
      \begin{minipage}[t]{0.315\hsize}
        \centering
        \includegraphics[keepaspectratio, scale=0.32]{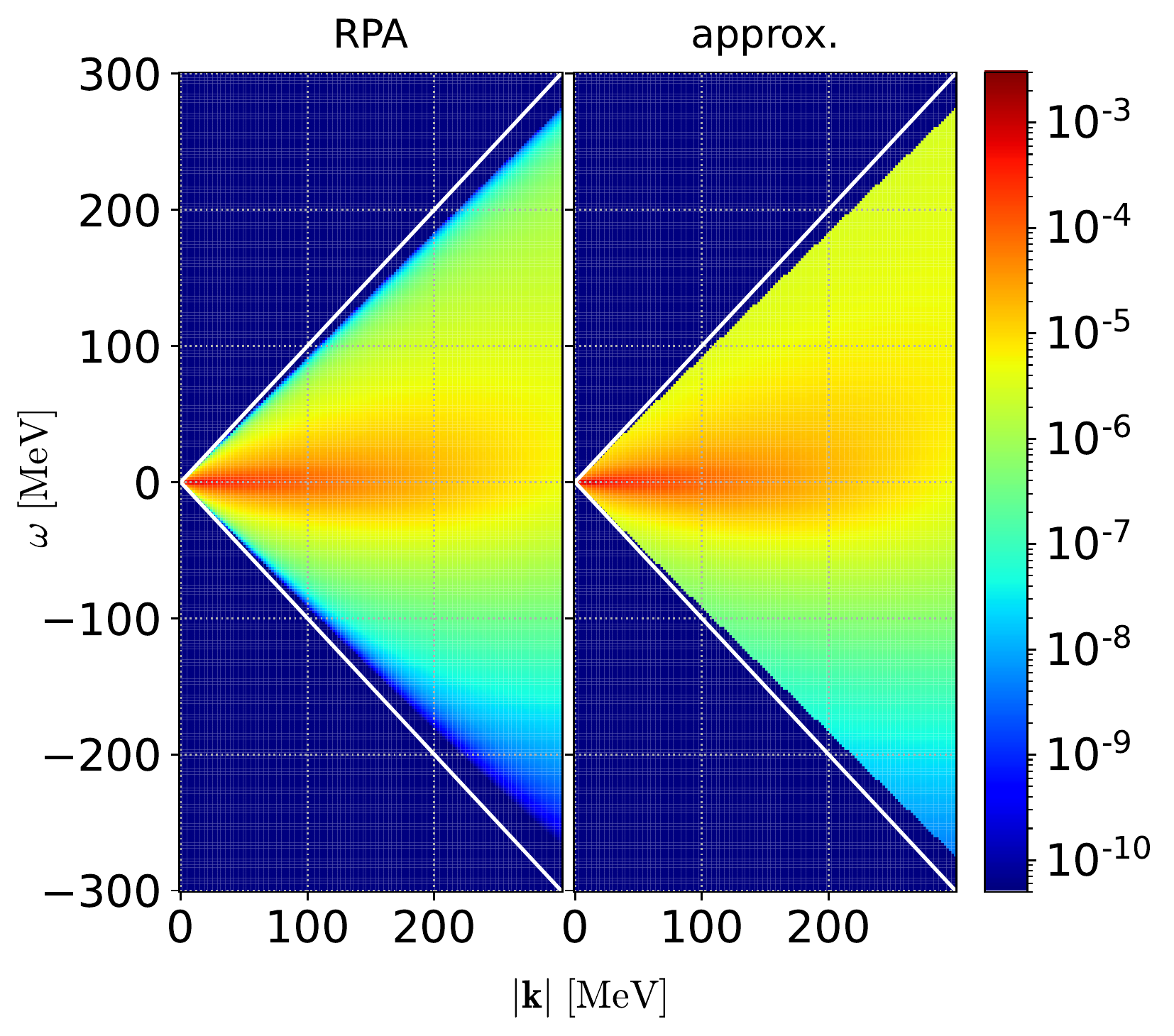}
        \subcaption{$T=0.9T_c$}
      \end{minipage} 
      \begin{minipage}[t]{0.315\hsize}
        \centering
        \includegraphics[keepaspectratio, scale=0.32]{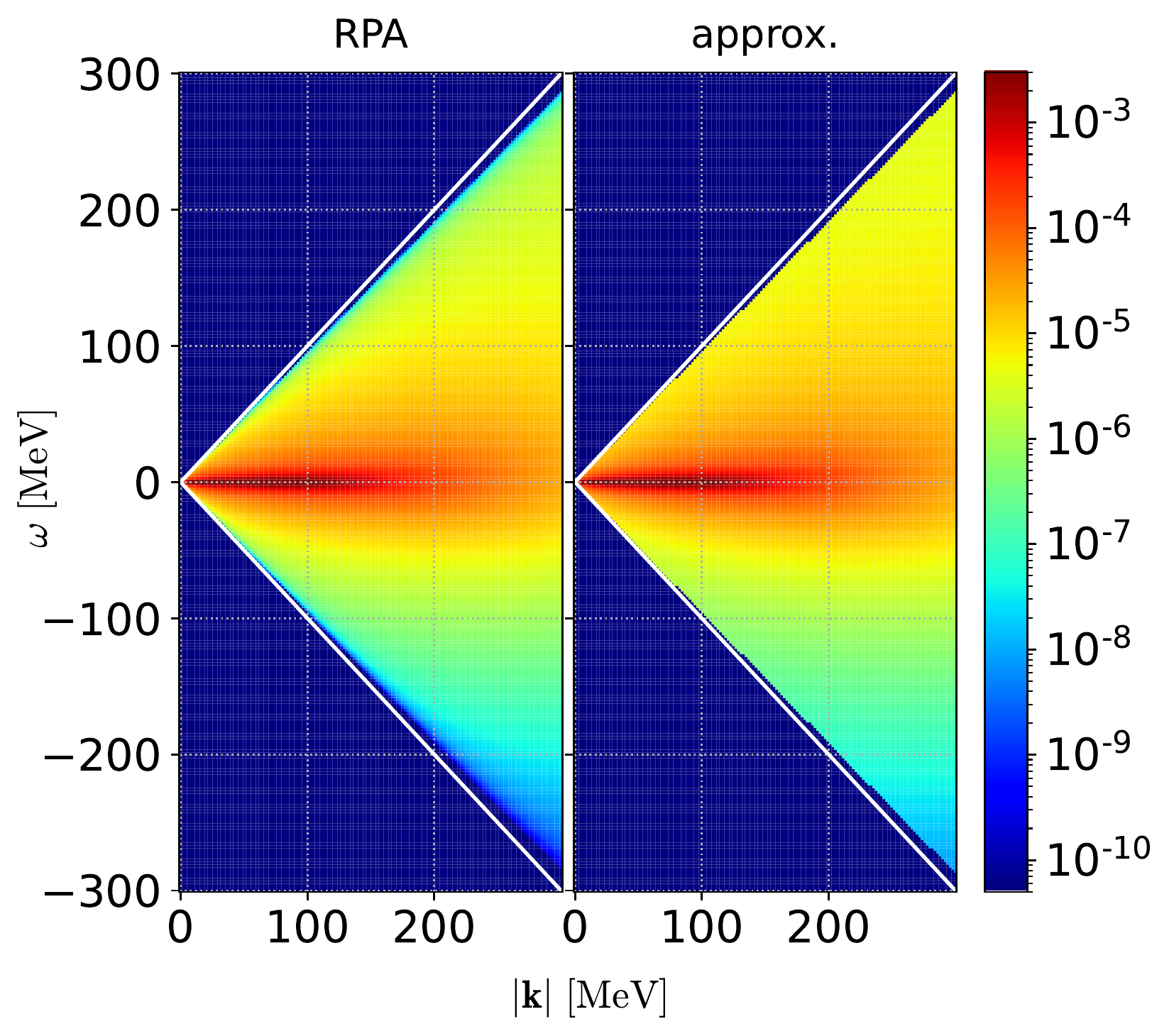}
        \subcaption{$T=1.0T_c$}
      \end{minipage} 
      \begin{minipage}[t]{0.315\hsize}
        \centering
        \includegraphics[keepaspectratio, scale=0.32]{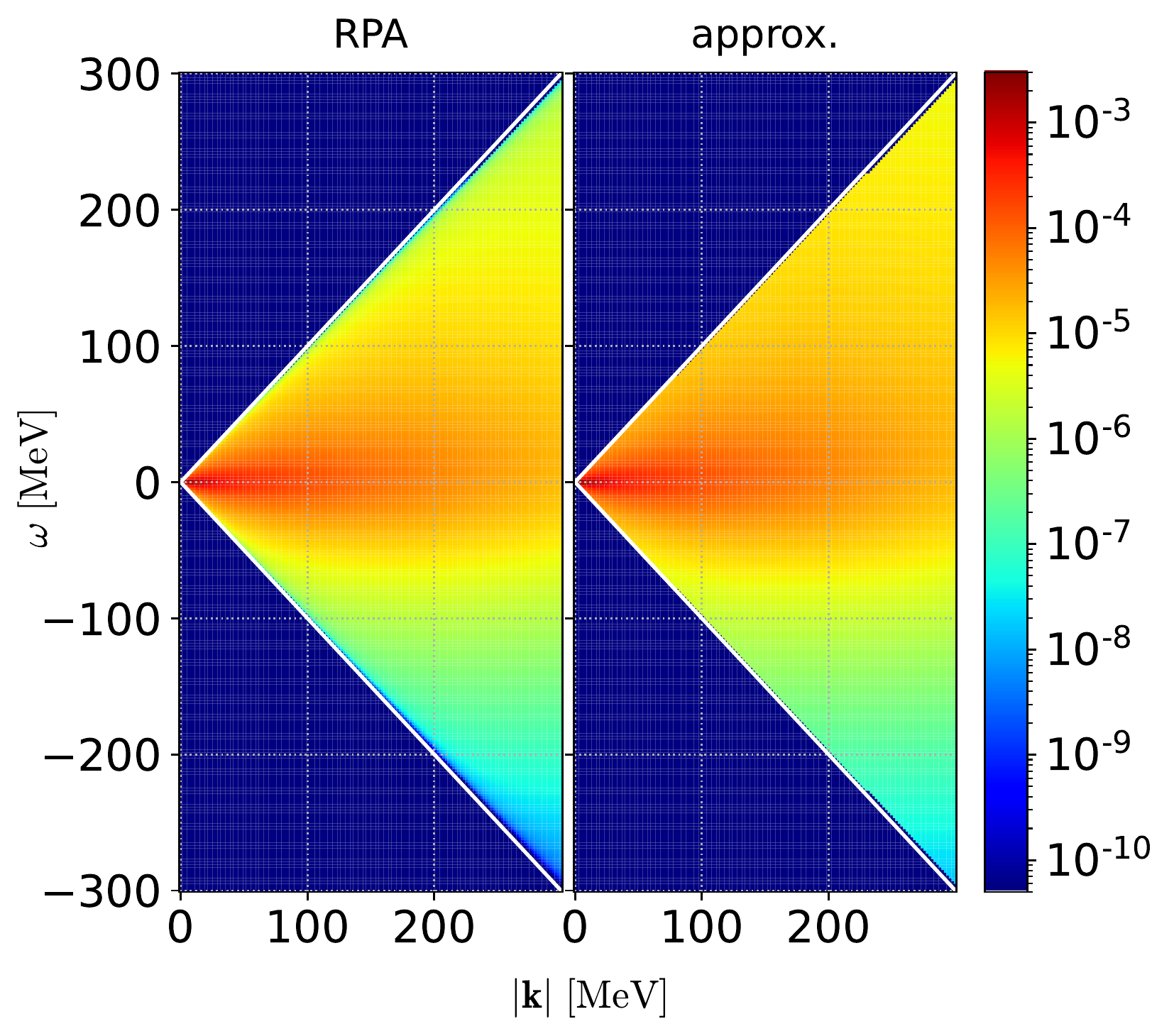}
        \subcaption{$T=1.1 T_c$}
      \end{minipage} 
     \end{tabular}
    \caption{
       Color maps of the dynamical structure factor 
       $S(\bm{k}, \omega)$ {in the space-like region}
       for $T=0.9T_c, 1.0T_c$ and $1.1T_c$, respectively.
       The solid (white) lines represent the light cone.
       The left subpanel for each $T$ is the result computed 
       by the RPA (\ref{eq:Xi}) and (\ref{eq:Q}),
       while the right one shows the approximate formula~(\ref{eq:TDGL-like}).}
     \label{fig:DSF}
\end{figure*}

To demonstrate the validity of Eq.~(\ref{eq:TDGL-like}),
we show in Fig.~\ref{fig:DSF} the contour maps 
of the dynamical structure factor given by
\begin{align}
S(\bm{k}, \omega) = \frac{1}{\pi} \frac{1}{1-e^{-\omega/T}}
{\rm Im} \Xi^R(\bm{k}, \omega) ,
\label{eq:DSF}
\end{align}
in the space-like region at and slightly away from the CP 
($T=0.9T_c, 1.0T_c$ and $1.1T_c$ at $\mu = \mu_c$).
In each panel, the left and right subpanels are the results of the RPA, 
Eq.~(\ref{eq:Xi}) with Eq.~(\ref{eq:ImQ}), and the approximation~(\ref{eq:TDGL-like}), respectively.
One finds that the former is well reproduced by the latter especially 
at the low energy region at which the soft mode has a significant strength.

Although the spectral properties of the soft mode of the QCD CP look quite similar 
to that of the diquark soft mode \cite{Nishimura:2022mku},
the present $\Xi^R(\bm{k},\omega)$ is not analytic 
at the origin $(|\bm{k}|,\omega)=(0,0)$ 
and it has a discontinuity at the light cone, 
in contrast to that of the diquark mode;
the discontinuity of the diquark propagator 
coming from the light cone 
is located at $|\omega+2\mu|=|\bm{k}|$, 
and accrodingly analytic at the origin~\cite{Kitazawa:2005vr,Nishimura:2022mku}.
This difference makes the following analysis 
require some extra caution in the present case.

\section{Dilepton production rate and electric conductivity}
\label{Photon self-energy}

In this section, we shall calculate the dilepton production rate (DPR) 
and the electric conductivity assuming that the system is
in the vicinity of the QCD CP.
In this case, these observables would be significantly modified by the soft modes.
Their effects are incorporated through
the calculation of the photon self-energy by taking them into account.
We perform this analysis in a parallel way to the analysis 
in Ref.~\cite{Nishimura:2022mku} that investigated the diquark soft modes.
Once the retarded photon self-energy $\Pi^{R \mu \nu} (\bm{k}, \omega)$ 
is obtained, the DPR is calculated to be
\begin{align}
\frac{d^4\Gamma}{d^4k}(\bm{k}, \omega) = - \frac{\alpha}{12\pi^4} 
\frac{1}{k^2} \frac{1}{e^{\omega/T}-1} g_{\mu\nu} {\rm Im} \Pi^{R \mu\nu} (\bm{k}, \omega),
\label{eq:DPR}
\end{align}
with the fine structure constant $\alpha$ and the Minkowski metric $g_{\mu \nu}$.
The electric conductivity $\sigma$ is also obtained as~\cite{book_Kapusta}
\begin{align}
& \ \ \sigma = \frac{1}{3} \lim_{\omega \rightarrow 0} \frac{1}{\omega} 
\sum_{i=1, 2, 3}  {\rm Im} \Pi^{R ii} (\bm{0}, \omega).
\label{eq:conductivity}
\end{align}

\subsection{Modification of photon self-energy by the soft mode}
\begin{figure*}[t]
    \begin{tabular}{cc}
      \begin{minipage}[t]{0.16\hsize}
        \centering
        \includegraphics[keepaspectratio, scale=0.09]{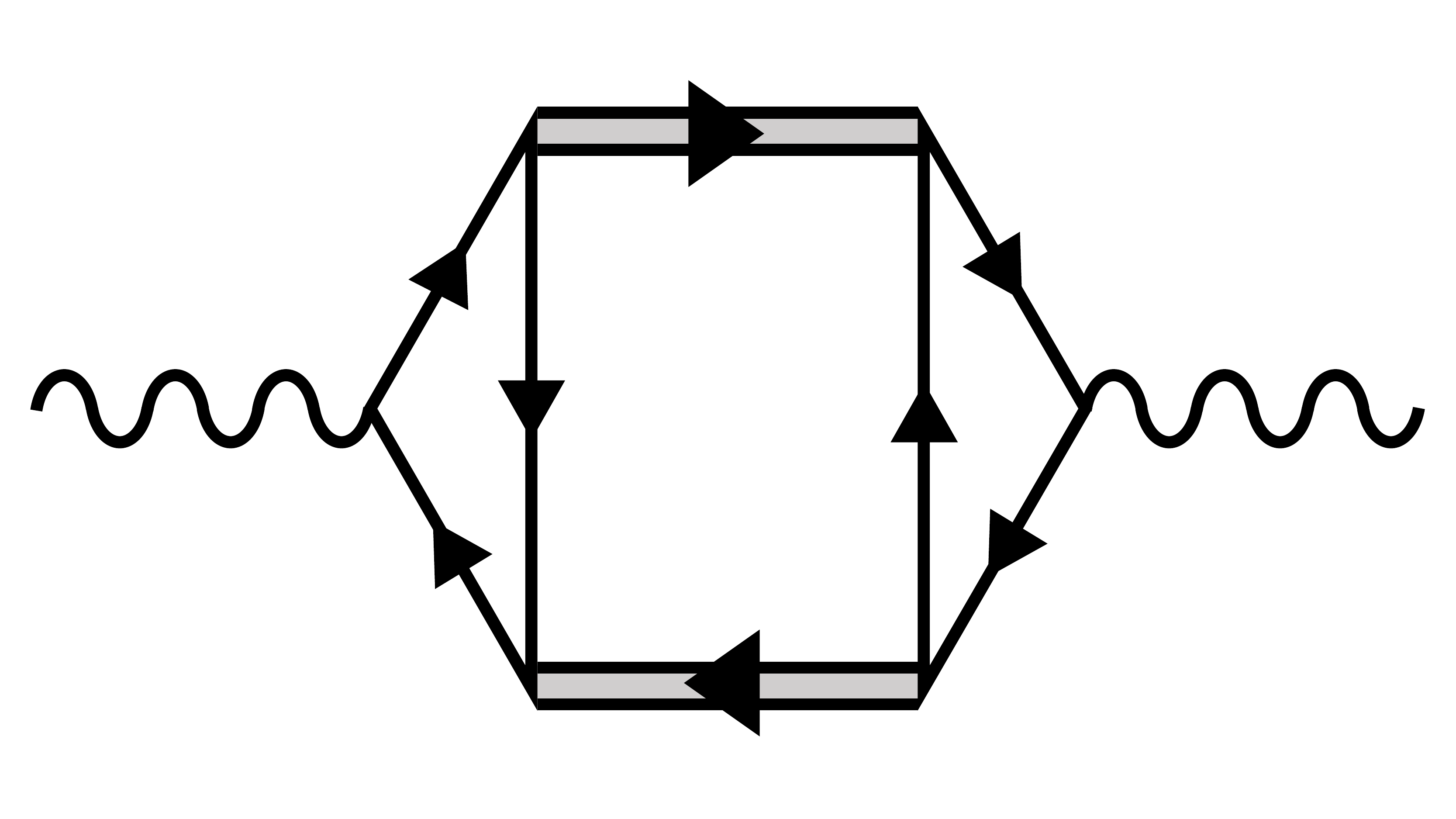}
        \subcaption{}
      \end{minipage} \\
      \begin{minipage}[t]{0.16\hsize}
        \centering
        \includegraphics[keepaspectratio, scale=0.09]{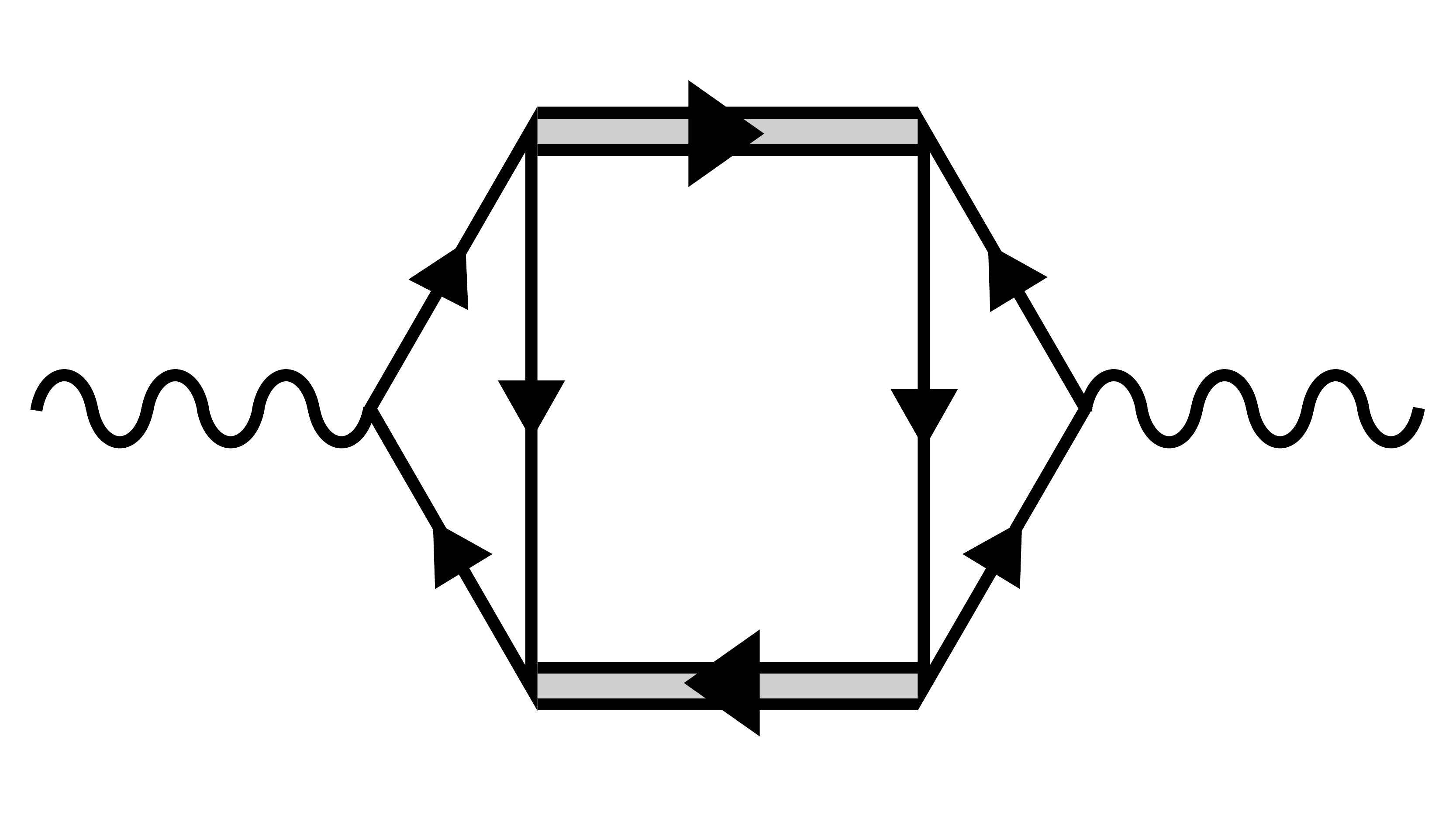}
        \subcaption{}
      \end{minipage} 
     \end{tabular}
     \begin{tabular}{cc}
      \begin{minipage}[t]{0.16\hsize}
        \centering
        \includegraphics[keepaspectratio, scale=0.09]{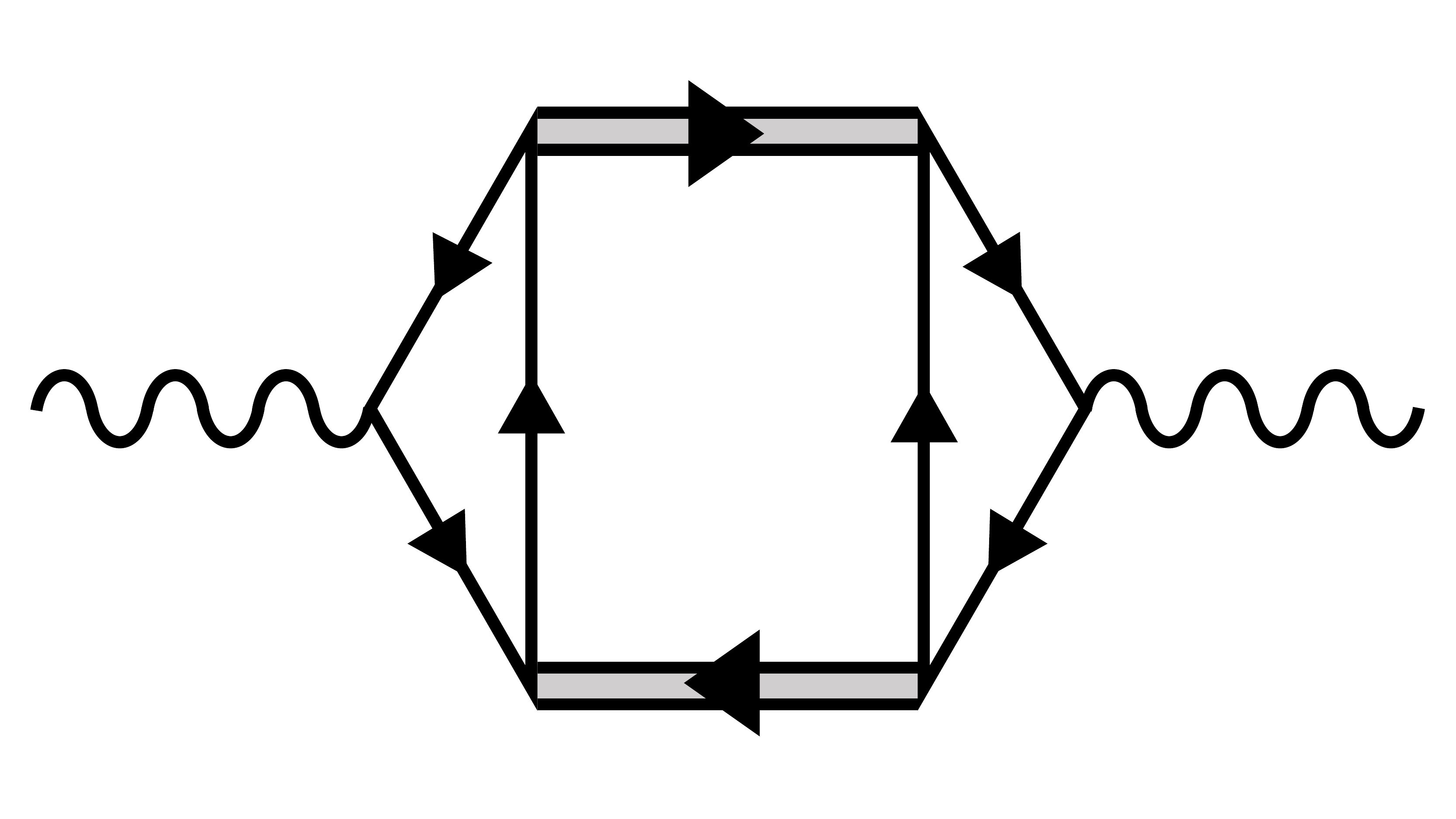}
        \subcaption{}
      \end{minipage} \\
      \begin{minipage}[t]{0.16\hsize}
        \centering
        \includegraphics[keepaspectratio, scale=0.09]{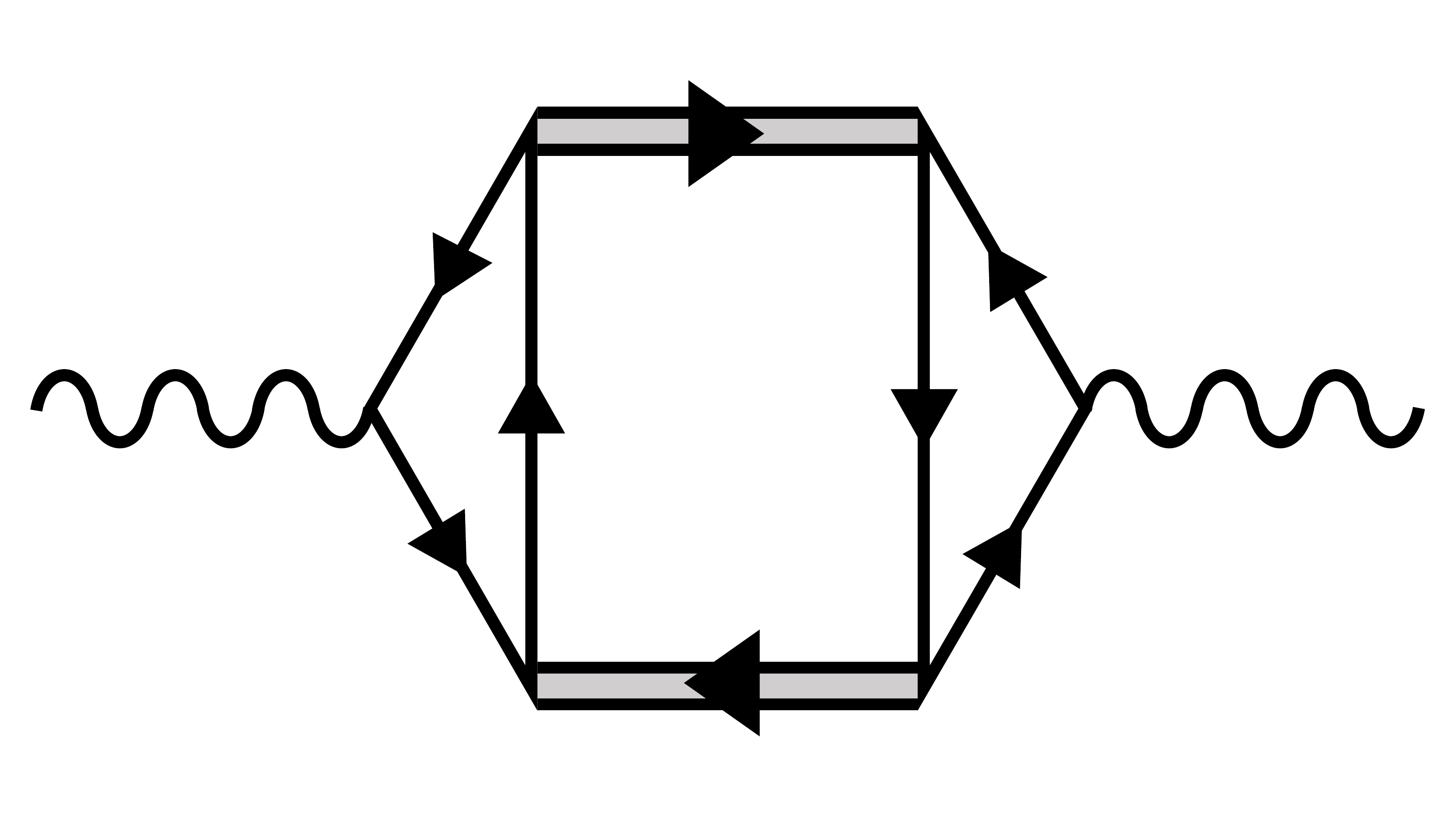}
        \subcaption{}
      \end{minipage} 
     \end{tabular}
     \begin{tabular}{cc}
      \begin{minipage}[t]{0.16\hsize}
        \centering
        \includegraphics[keepaspectratio, scale=0.09]{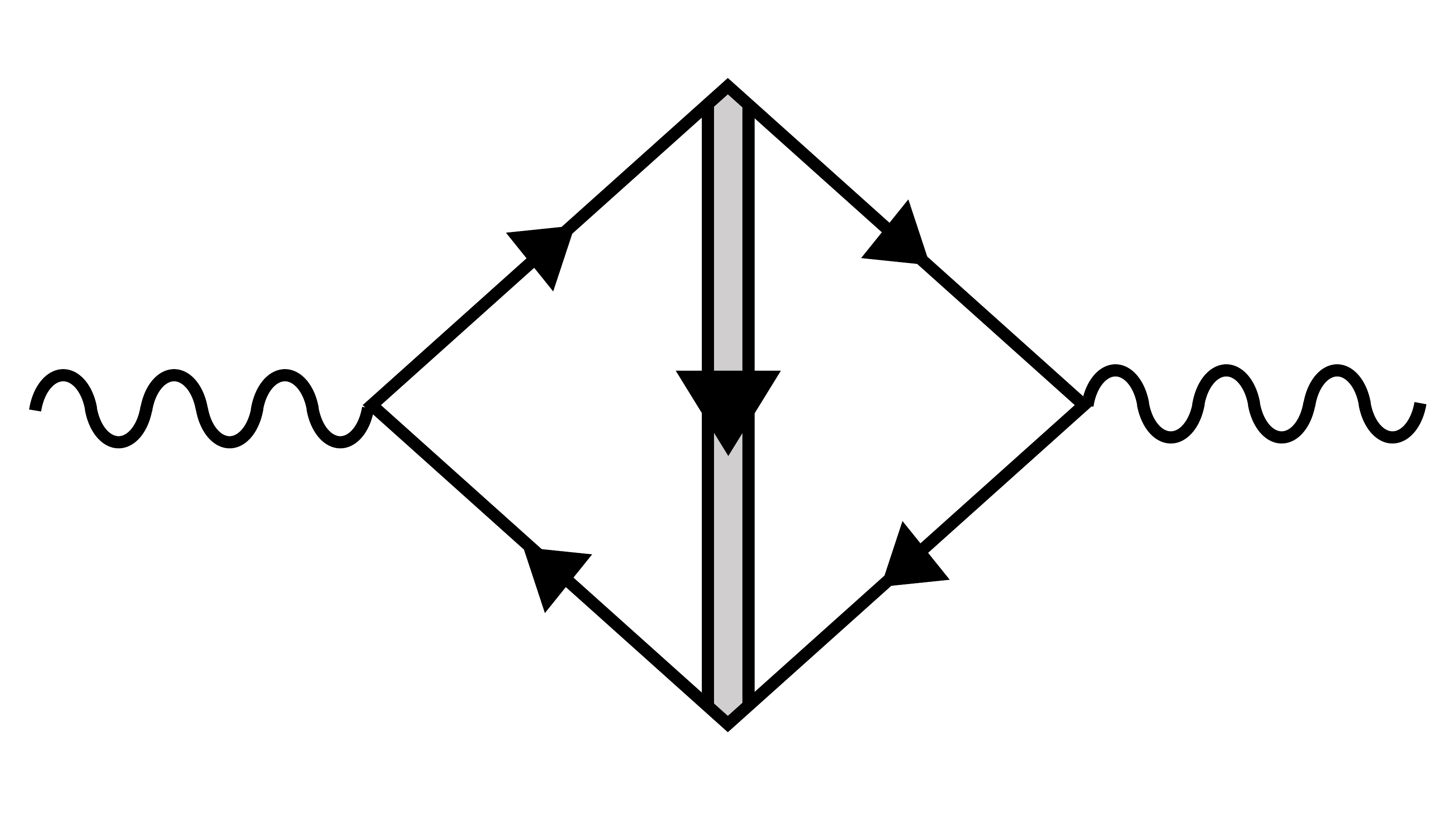}
        \subcaption{}
      \end{minipage} \\
      \begin{minipage}[t]{0.16\hsize}
        \centering
        \includegraphics[keepaspectratio, scale=0.09]{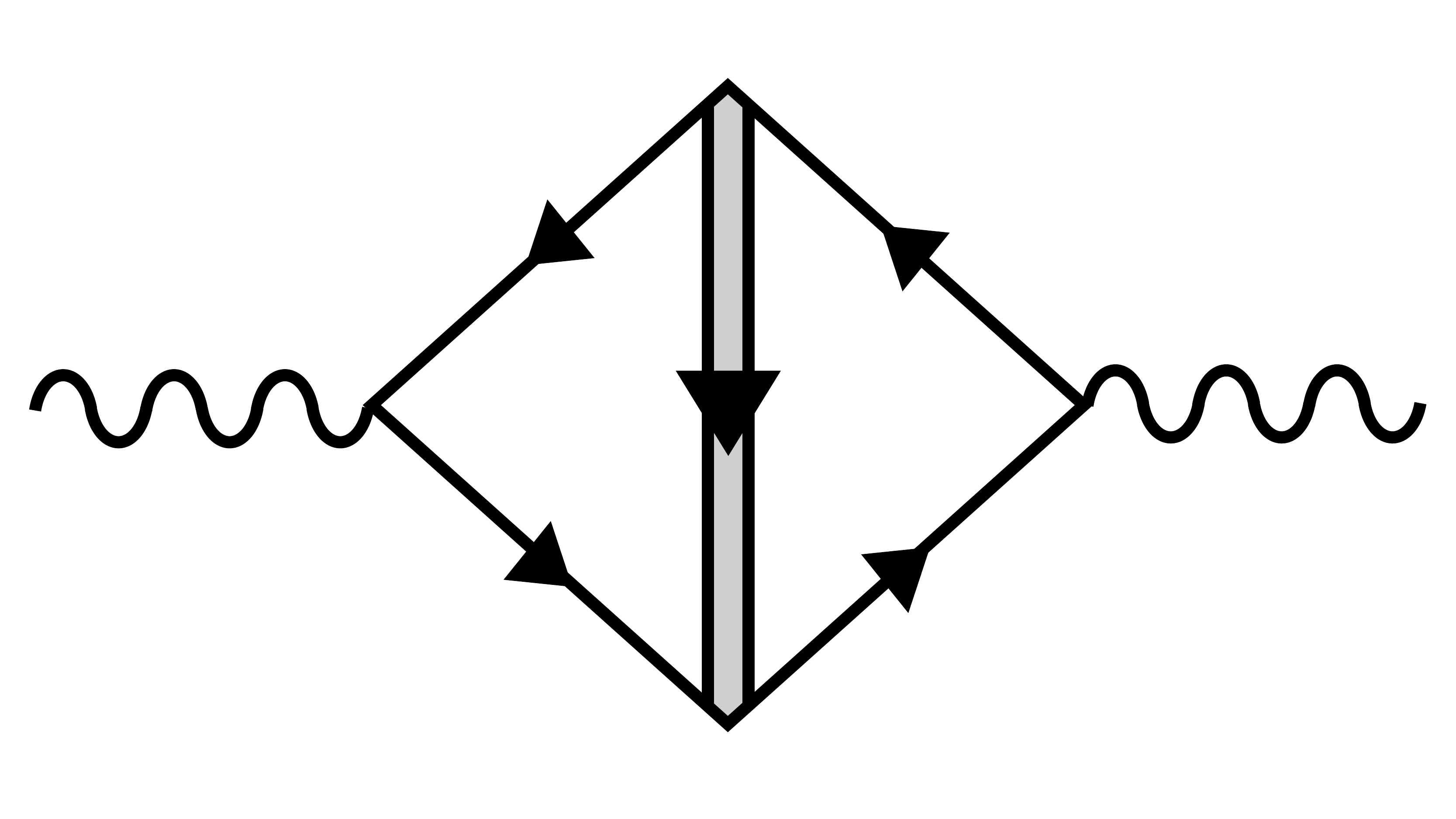}
        \subcaption{}
      \end{minipage} 
     \end{tabular}
     \begin{tabular}{cc}
      \begin{minipage}[t]{0.16\hsize}
        \centering
        \includegraphics[keepaspectratio, scale=0.09]{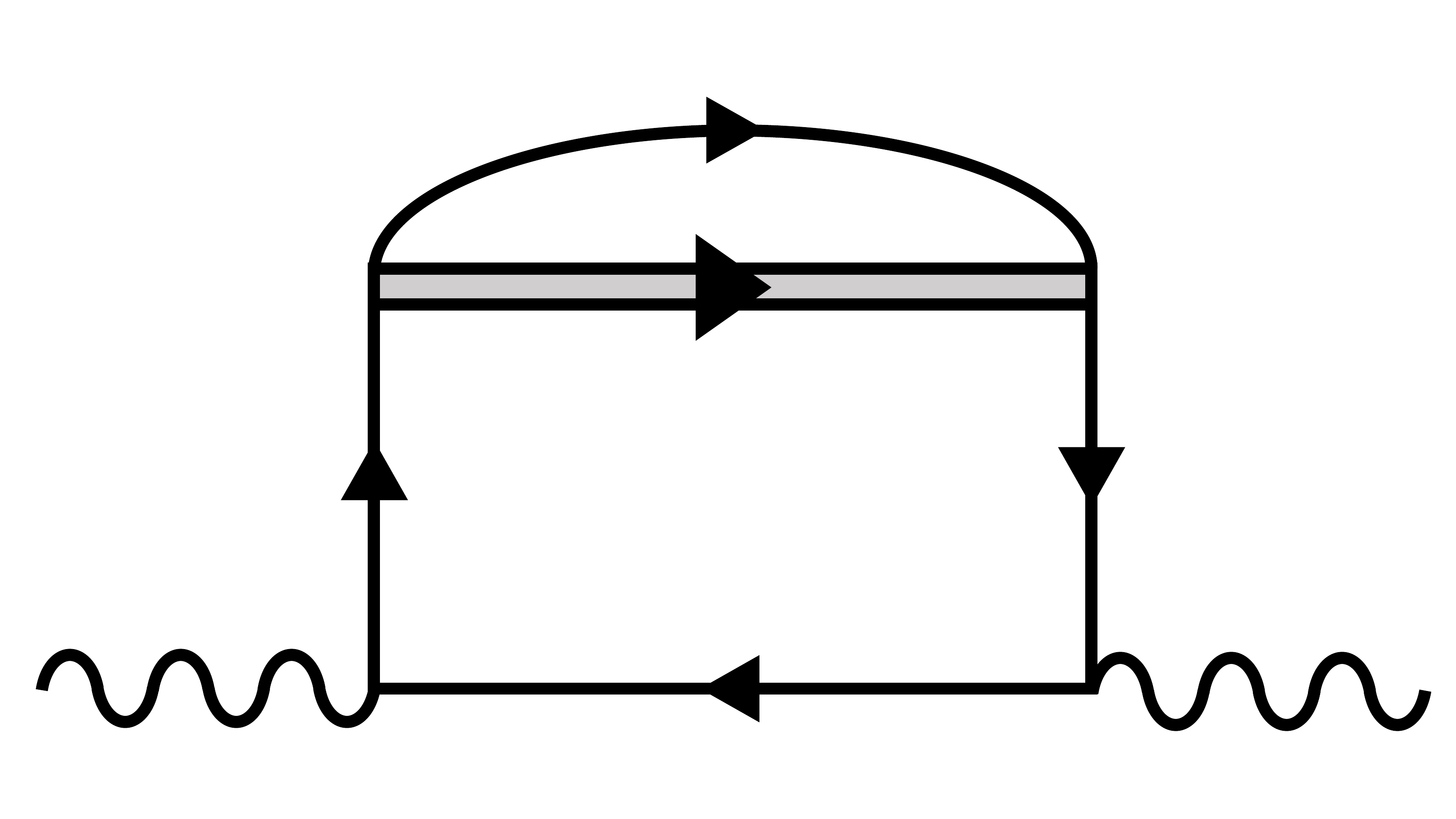}
        \subcaption{}
      \end{minipage} \\
      \begin{minipage}[t]{0.16\hsize}
        \centering
        \includegraphics[keepaspectratio, scale=0.09]{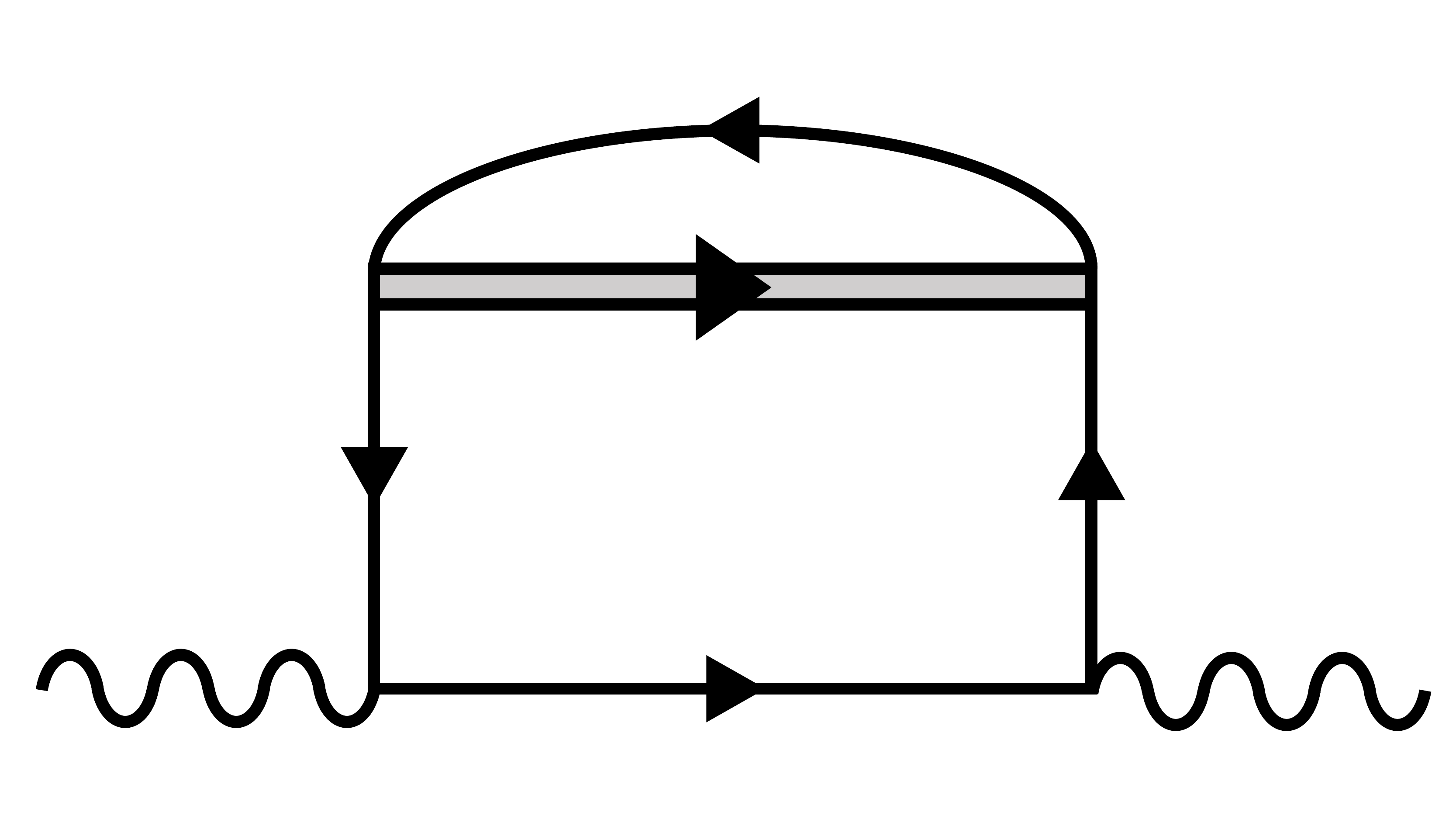}
        \subcaption{}
      \end{minipage} 
    \end{tabular}
    \begin{tabular}{cc}
      \begin{minipage}[t]{0.16\hsize}
        \centering
        \includegraphics[keepaspectratio, scale=0.09]{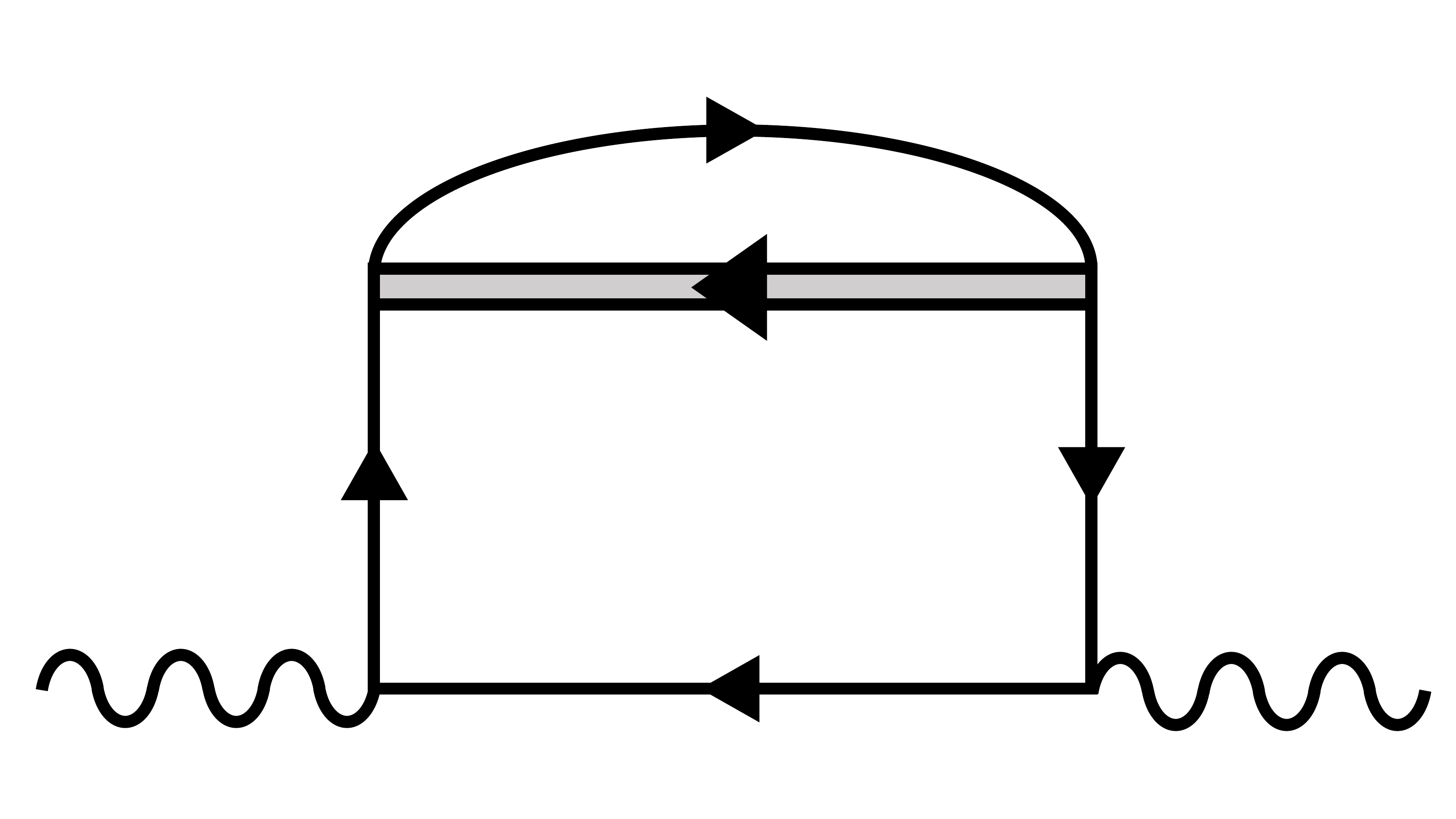}
        \subcaption{}
      \end{minipage} \\
      \begin{minipage}[t]{0.16\hsize}
        \centering
        \includegraphics[keepaspectratio, scale=0.09]{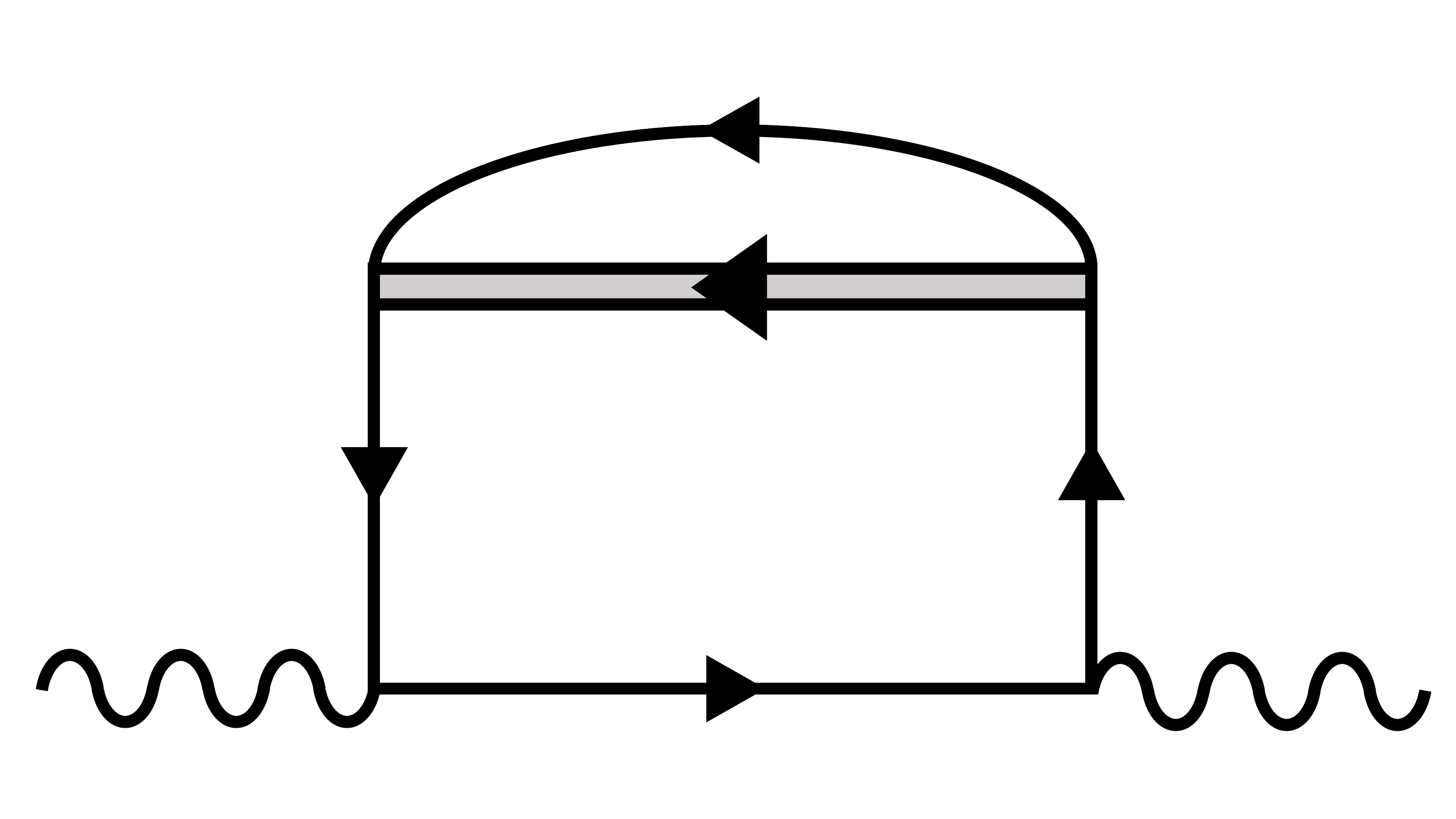}
        \subcaption{}
      \end{minipage} 
     \end{tabular}
    \caption{
Diagrammatic representations of the Aslamazov-Larkin (a)--(d),
Maki-Thompson (e, f) and density of states (g)--(j) terms with the soft modes.
The single, double and wavy lines are quarks, 
soft modes and photons, respectively.
}
\label{fig:self-energy}
\end{figure*}

To construct the photon self-energy in a gauge-invariant way,
we start from the lowest-order contribution of the soft mode
to the thermodynamic potential 
$\Omega_{\rm fluc} = \int_p \ln [G_S \tilde\Xi^{-1}(p)]$,
which is diagrammatically represented by the one-loop graph 
of the soft mode propagator.
The photon self-energy is then constructed by attaching
electromagnetic vertices at two points of quark lines in $\Omega_{\rm fluc}$.
This procedure leads to ten types of diagrams shown in Fig.~\ref{fig:self-energy}.
By borrowing the nomenclature in the theory 
of superconductivity~\cite{AL:1968,Maki:1968,Nishimura:2022mku}, 
we call (a)--(d) the Aslamazov-Larkin (AL) \cite{AL:1968}, 
(e) and (f) the Maki-Thompson (MT) \cite{Maki:1968} 
and (g)--(j) the density of states (DOS) terms, respectively.
The respective contributions to the photon self-energy, 
$\tilde{\Pi}^{\mu \nu}_{\rm AL} (k)$, $\tilde{\Pi}^{\mu \nu}_{\rm MT} (k)$ 
and $\tilde{\Pi}^{\mu \nu}_{\rm DOS} (k)$,
in the imaginary-time formalism are expressed as
\begin{align}
\Pi_{\rm AL}^{\mu\nu} (k) &= \sum_f \int_q 
\tilde{\Gamma}^\mu_f(q, q+k) \tilde{\Xi}(q+k) 
\tilde{\Gamma}^\nu_f(q+k, q) \tilde{\Xi}(q), 
\label{eq:AL} \\
\Pi_{\rm MT}^{\mu\nu} (k) &= \sum_f \int_q 
\tilde{\Xi}(q) \mathcal{R}_{{\rm MT},f}^{\mu\nu} (q, k), 
\label{eq:MT} \\
\Pi_{\rm DOS}^{\mu\nu} (k) &= \sum_f \int_q 
\tilde{\Xi}(q) \mathcal{R}_{{\rm DOS},f}^{\mu\nu} (q, k), 
\label{eq:DOS} 
\end{align}
where $f=u,d$ is the index of flavors and
the vertex functions $\tilde{\Gamma}^\mu_f(q, q+k)$, 
$\mathcal{R}_{{\rm MT},f}^{\mu\nu}(q, k)$ and 
$\mathcal{R}_{{\rm DOS},f}^{\mu\nu}(q, k)$ 
represent the three- and four-point diagrams in Fig.~\ref{fig:self-energy}.
We note that the number of diagrams is doubled compared 
with the case in Ref.~\cite{Nishimura:2022mku}, 
since the quark and anti-quark lines should be distinguished for the present case.

The total photon self-energy reads
\begin{align}
\tilde{\Pi}^{\mu\nu} (k)&= \tilde{\Pi}_{\rm free}^{\mu\nu} (k) 
+ \tilde{\Pi}_{\rm fluc}^{\mu\nu} (k), \\
\tilde{\Pi}_{\rm fluc}^{\mu\nu} (k) &= \tilde{\Pi}_{\rm AL}^{\mu\nu} (k) 
+ \tilde{\Pi}_{\rm MT}^{\mu\nu} (k) + \tilde{\Pi}_{\rm DOS}^{\mu\nu} (k),
\end{align}
where $\tilde{\Pi}_{\rm fluc}^{\mu\nu} (k)$ 
is the contribution from the soft modes and 
\begin{align}
\tilde{\Pi}_{\rm free}^{\mu\nu} (k) = N_c C_{\rm em} 
\int_p {\rm Tr}[\gamma^\mu {\cal G}(p+k) \gamma^\nu {\cal G}(p)],
\end{align}
is the self-energy of the free-quark system,
where $C_{\rm em} = e_u^2+e_d^2$ 
with $e_u=2|e|/3$ ($e_d=-|e|/3$) 
denoting the electric charges of up (down) quark.
We note that $\tilde{\Pi}^{\mu\nu} (k)$ thus constructed 
nicely satisfies the Ward-Takahashi (WT) identity 
\begin{align}
k_\mu \tilde{\Pi}^{\mu\nu} (k) = 0 .
\label{eq:self-energy-WT} 
\end{align}

\subsection{Vertices}
For the vertex functions $\tilde{\Gamma}^\mu_f(q, q+k)$, 
$\mathcal{R}_{{\rm MT},f}^{\mu\nu}(q, k)$ and 
$\mathcal{R}_{{\rm DOS},f}^{\mu\nu}(q, k)$, 
instead of calculating the diagrams in Fig.~\ref{fig:self-energy} directly 
we determine their functional forms from the WT identities for the vertices
\begin{align}
k_\mu \tilde{\Gamma}^\mu_f (q, q+k) &
= - e_f [\tilde{\Xi}^{-1}(q+k)-\tilde{\Xi}^{-1}(q)], 
\label{eq:AL-vertex-WT} \\
k_\mu \mathcal{R}^{\mu\nu}_f (q, k) &
= - e_f [\tilde{\Gamma}^\nu_f (q-k, q)-\tilde{\Gamma}^\nu_f (q, q+k)], 
\label{eq:MTDOS-vertex-WT}
\end{align}
where $\mathcal{R}^{\mu\nu}(q, k)=
\mathcal{R}_{{\rm MT},f}^{\mu\nu}(q, k)+\mathcal{R}_{{\rm DOS},f}^{\mu\nu}(q, k)$.

Among the vertex functions, only their spatial components are needed
for the calculations of Eqs.~(\ref{eq:DPR}) and~(\ref{eq:conductivity}),
because $\tilde{\Pi}_{\rm fluc}^{00}(k)$ in Eq.~(\ref{eq:DPR}) is obtained from the spatial components through
\begin{align}
  \tilde{\Pi}^{00} (k)
  = \frac{\bm{k}^2}{(i\nu_l)^2} \tilde{\Pi}^{11} (k)
  \qquad \mbox{for} \quad k=(|\bm{k}|, 0, 0, i\nu_l).
  \label{eq:Pi00}
\end{align}

To obtain $\tilde{\Gamma}^i_f (q, q+k)$ for $i=1,2,3$, 
we take the same procedure as that adopted in
Ref.~\cite{Nishimura:2022mku},
where the energy dependent and independent terms of $\tilde\Xi^{-1}(q)$ 
on the right-hand side in Eq.~(\ref{eq:AL-vertex-WT}) are attributed to 
$k_0\tilde{\Gamma}^0_f (q, q+k)$ 
and $\bm{k}\cdot\tilde{\bm\Gamma}_f (q, q+k)$ on the left-hand side,
respectively, 
so that the spatial part of Eq.~(\ref{eq:AL-vertex-WT}) 
is given by~\footnote{This procedure is justified by calculating $\tilde\Gamma^\mu_f(q,q+k)$ from the triangle diagrams in Fig.~\ref{fig:self-energy} (a)--(d) directly and comparing the functional forms in the small $\omega$ and $\bm{k}$ limit~\cite{Nishimura_prep}.}
\begin{align}
  \bm{k} \cdot \tilde{\bm{\Gamma}}_f (q, q+k)
  = e_f [A(\bm{q}+\bm{k})-A(\bm{q})].
  \label{eq:Gamma^i}
\end{align}
We then employ the ansatz on the form
of $\tilde{\Gamma}^i_f (q, q+k)$ that satisfies Eq.~(\ref{eq:Gamma^i}) as 
\begin{align}
\tilde{\Gamma}^i_f (q, q+k) &= e_f Q_{(1)}(\bm{q}+\bm{k}, \bm{q}) (2q+k)^i, 
\label{eq:AL-vertex-approx} \\
Q_{(1)}(\bm{q}_1, \bm{q}_2)
&=\frac{A(\bm{q}_1) - A(\bm{q}_2)}{|\bm{q}_1|^2 - |\bm{q}_2|^2}.
\end{align}
Since $A(\bm{q})$ is real, $\tilde{\Gamma}^i_f (q, q+k)$ 
is also a real function in this construction.
We note that this form of 
approximation is valid only for sufficiently small $k$, 
as the WT identity~(\ref{eq:AL-vertex-WT}) cannot 
uniquely determine the vertex in general.
Near the QCD CP at which the contribution of the soft mode becomes prominent,
 it is, however, expected that the qualitative result does not 
depend on the form of the vertex and our approximation would be well justified.

The form of the vertex $\mathcal{R}_{f}^{ij}(q, k)$ is also obtained by 
adopting a similar argument with Eqs.~(\ref{eq:MTDOS-vertex-WT}) 
and~(\ref{eq:AL-vertex-approx}), as was done in Ref.~\cite{Nishimura:2022mku}.
From this analysis it is found that $\mathcal{R}_{f}^{ij}(q, k)$ 
is a real function and independent of $i\nu_l$.
By constructing the MT and DOS terms from the vertex, one finds
\begin{align}
      {\rm Im}\Pi_{\rm MT}^{R ij} (\bm{k},\omega)
    +{\rm Im}\Pi_{\rm DOS}^{R ij} (\bm{k},\omega)=0.
    \label{eq:Im=0}
\end{align}
Equation~(\ref{eq:Im=0}) is shown from the fact that the sum of
Eqs.~(\ref{eq:MT}) and~(\ref{eq:DOS}) becomes real 
after the Matsubara summations when $\mathcal{R}_{f}^{ij}(q, k)$ satisfies 
the above conditions~\cite{Nishimura:2022mku,Nishimura_prep}.
The cancellation of the MT and DOS terms is also known 
in metallic superconductivity~\cite{book_Larkin}.

From Eqs.~(\ref{eq:Im=0}) and~(\ref{eq:Pi00}),
one obtains
\begin{align}
  {\rm Im}\Pi_{\rm fluc}^{R00}(\bm{k},\omega)
  = \frac{\bm{k}^2}{\omega^2}
  {\rm Im}\Pi_{\rm AL}^{R11}(\bm{k},\omega)
  \qquad
  \mbox{for}
  \qquad
  \bm{k}=(|\bm{k}|,0,0).
  \label{eq:ImPIflucAL}
\end{align}
Plugging this into Eq.~(\ref{eq:DPR}),
one finds that the DPR is written 
solely in terms of the AL term.
So is the electric conductivity
since it is given by the spatial components of 
${\rm Im}\Pi_{\rm fluc}^{R\mu\nu}(\bm{k},\omega)$ 
as in Eq.~(\ref{eq:conductivity}).
These results show that we only have to compute the AL term
for obtaining both the DPR and the electric conductivity.

\subsection{Aslamazov-Larkin term}
Since ${\rm Im}\tilde{\Pi}_{\rm fluc}^{ij}(k)$ consists of only  the AL term, 
we now calculate $\tilde{\Pi}_{\rm AL}^{ij}(k)$.
Using Eqs.~(\ref{eq:TDGL-like}) and (\ref{eq:AL-vertex-approx}),
we obtain 
\begin{align}
\tilde{\Pi}^{ij}_{\rm AL} (k)
=&\sum_f \int \frac{d^3\bm{q}}{(2\pi)^3}
\tilde{\Gamma}^i_f (q, q+k) \tilde{\Gamma}^j_f (q+k, q) 
\oint_C \frac{dq_0}{2\pi i} \frac{{\rm coth} \frac{q_0}{2T}}{2} 
\tilde{\Xi} (q+k) \tilde{\Xi} (q),
\label{eq:Pi_fluc}
\end{align}
where the contour $C$ encircles the imaginary axis,
which is  deformed so as to avoid
the cut in $\tilde{\Xi} (q+k)$ and $\tilde{\Xi} (q)$.

Taking the analytic continuation $i\nu_l \rightarrow \omega + i\eta$ 
and using Eq.~(\ref{eq:ImPIflucAL}) we obtain
\begin{align}
g_{\mu \nu} {\rm Im} \Pi^{R \mu \nu}_{\rm fluc} (\bm{k}, \omega)
=&
\frac{\bm{k}^2}{\omega^2} {\rm Im} \Pi^{R 11}_{\rm AL} (\bm{k}, \omega)
- \sum_i {\rm Im} \Pi^{R ii}_{\rm AL} (\bm{k}, \omega)
\nonumber \\
=&C_{\rm em}
\int \frac{d^3\bm{q}}{(2\pi)^3} 
\int \frac{d\omega'}{2\pi} {\rm coth} \frac{\omega'}{2T}
\nonumber \\
&\times \bigl( Q_{(1)}(\bm{q}+\bm{k}, \bm{q}) \bigr)^2
\biggl[
\biggl( \frac{(\bm{q}+\bm{k})^2-\bm{q}^2}{\omega} \biggr)^2 - (2\bm{q}+\bm{k})^2
\biggr]
\nonumber \\
&\times {\rm Im} \Xi^R(\bm{q}+\bm{k}, \omega')
\bigl\{
{\rm Im} \Xi^R(\bm{q}, \omega'+\omega) - {\rm Im} \Xi^R(\bm{q}, \omega'-\omega)
\bigr\}.
\label{eq:ImPi_fluc}
\end{align}
The contribution of the soft mode to the DPR is computed 
by substituting Eq.~(\ref{eq:ImPi_fluc}) into Eq.~(\ref{eq:DPR}).
The contribution of the soft mode to the electric conductivity is
also obtained by plugging the formula
\begin{align}
\sum_{i=1}^3 {\rm Im} \Pi^{R ii}_{\rm fluc} (\bm{k}, \omega)
=&C_{\rm em}
\int \frac{d^3\bm{q}}{(2\pi)^3} 
\int \frac{d\omega'}{2\pi} {\rm coth} \frac{\omega'}{2T}
\bigl( Q_{(1)}(\bm{q}+\bm{k}, \bm{q}) \bigr)^2 (2\bm{q}+\bm{k})^2
\nonumber \\
&\times {\rm Im} \Xi^R(\bm{q}+\bm{k}, \omega')
\bigl\{
{\rm Im} \Xi^R(\bm{q}, \omega'+\omega) - {\rm Im} \Xi^R(\bm{q}, \omega'-\omega)
\bigr\},
\label{eq:ImPi_fluc2}
\end{align}
into Eq.~(\ref{eq:conductivity}).
We note that Eq.~(\ref{eq:ImPi_fluc2}) 
at $|\bm{k}|=0$ is linearly dependent on $\omega$
in the $\omega\to0$ limit, and hence the conductivity $\sigma$ 
calculated from it has a nonzero value. 
This term leads to the divergence of $\sigma$ at the QCD CP 
as we will see in the next section.

We note that the domain of the integral 
in Eq.~(\ref{eq:ImPi_fluc}) or~(\ref{eq:ImPi_fluc2}) is subject to a constraint 
that Eq.~(\ref{eq:TDGL-like}) takes a nonzero value 
only in the energy-momentum region $|\omega|<\bar{k} (|\bm{k}|, \bar{\Lambda})$, 
i.e., inside the space-like region.
Nevertheless, we note that some multiple soft mode processes 
can affect the photon self-energy in the time-like region 
that is responsible for the DPR and conductivity.
These contributions are understood as the scattering process 
of the photon with a soft mode:
Let a virtual photon with the energy-momentum 
$k=(\bm{k}, \omega)$ is absorbed by a soft mode with $q_1=(\bm{q}_1, \omega_1)$ 
to make another one with $q_2=(\bm{q}_2, \omega_2)$,
both of which are in the space-like region; 
$|\omega_1|<|\bm{q}_1|$ and $|\omega_2|<|\bm{q}_2|$.
Then, the energy-momentum conservation law tells us that
$\bm{k}=\bm{q}_2-\bm{q}_1$ and $\omega=\omega_2-\omega_1$,
where $|\bm{k}|$ can be taken arbitrarily small
keeping $\omega=\omega_2-\omega_1$ finite.
Thus, the soft mode which has the spectral support in the space-like region
can contribute to the photon self-energy in the time-like region ($\omega>|\bm{k}|$).
In the next section we shall see that it can cause 
an enhancement of the DPR and conductivity.

Before closing this section, let us clarify the limitation of our calculation. 
Firstly, in our treatment we focus on the effects 
of the soft mode in the space-like region, and the effects 
of the mesonic mode in the time-like region is neglected. 
This approximation is justified as long as we consider the DPR 
in the low energy region near the QCD CP, 
since the mass of the sigma mode is larger than $2M\simeq370$~MeV. 
When considering the DPR above $2M$, however, 
the effect of the mesonic modes will become significant.
Secondly, we have constructed the approximate forms of
the vertex functions through the WT identities
and Eq.~(\ref{eq:XiAB}).
While this assumption should be valid for sufficiently small $\omega$, 
it would not be directly applicable to the large energy-momentum region.

\section{Numerical results}
\label{Numerical results}
\begin{figure*}[t]
    \begin{tabular}{cc}
      \begin{minipage}[t]{0.48\hsize}
        \centering
        \includegraphics[keepaspectratio, scale=0.4]{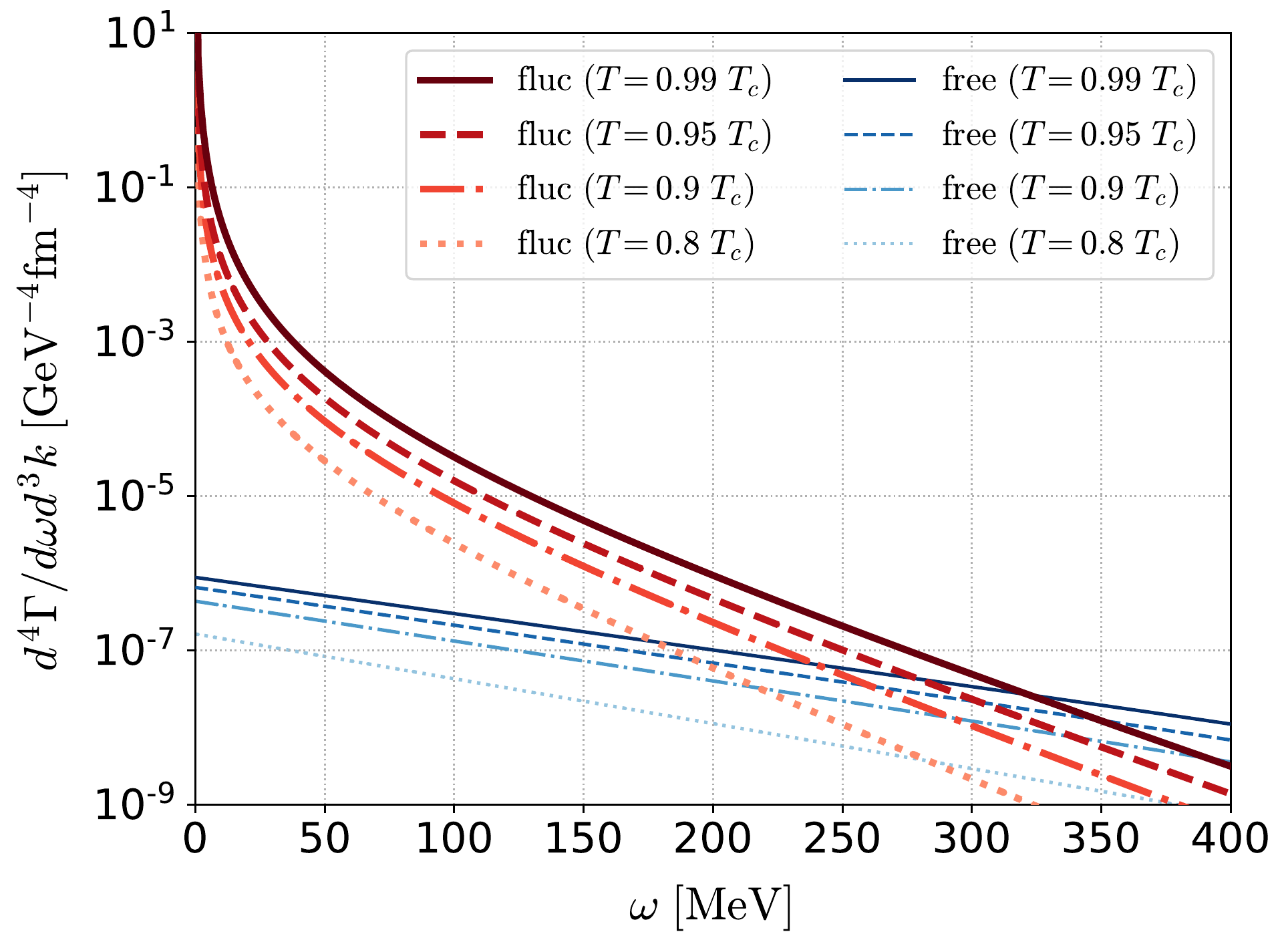}
      \end{minipage} 
      \begin{minipage}[t]{0.48\hsize}
        \centering
        \includegraphics[keepaspectratio, scale=0.4]{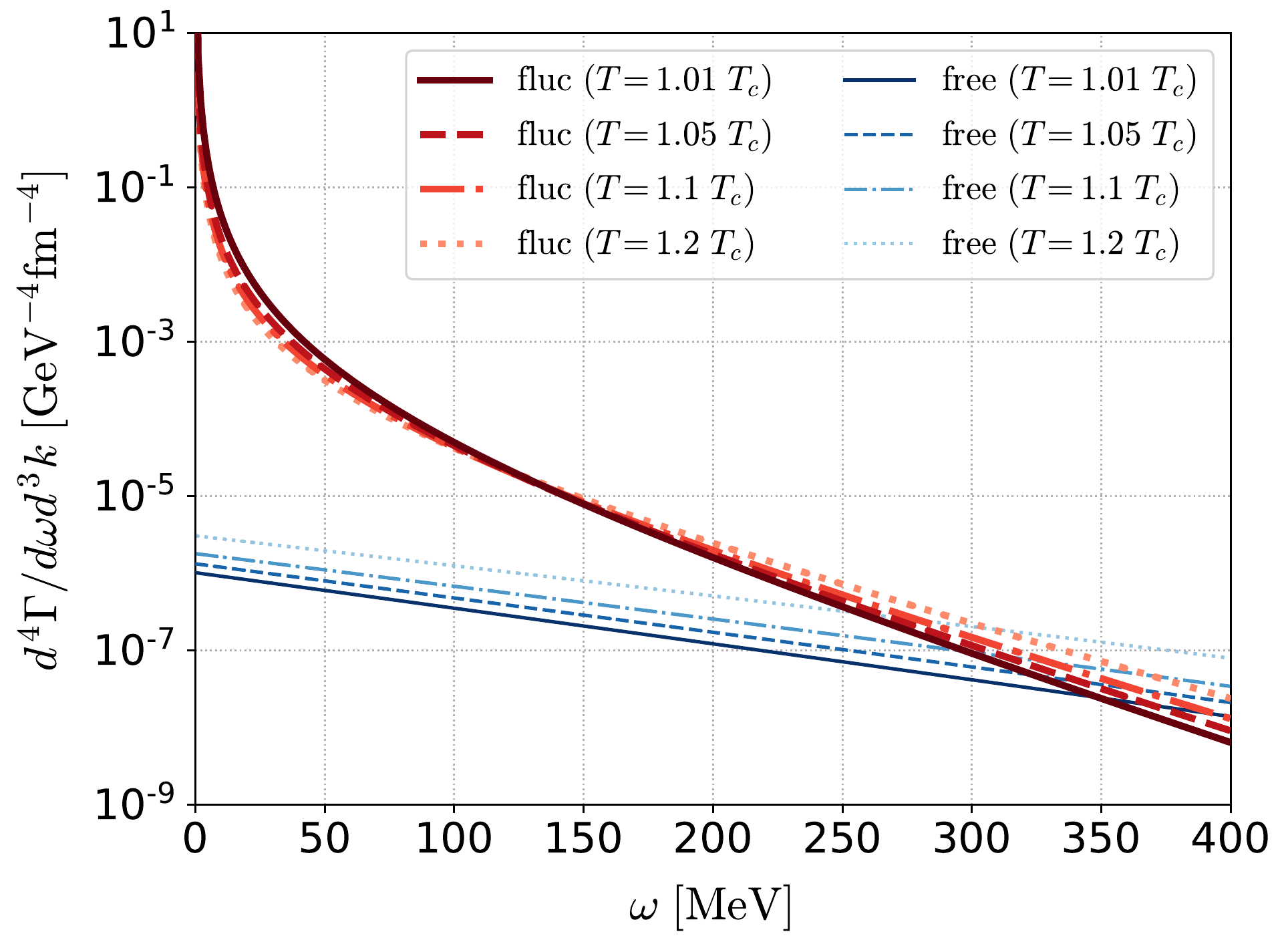}
      \end{minipage} 
     \end{tabular}
\caption{
Dilepton production rate (DPR) per unit energy and momentum
$d^4\Gamma / d\omega d^3k$ for several values of $T/T_c$ at $\mu=\mu_c$ 
and $\bm{k}=\bm{0}$.
The thick (red) and thin (blue) lines are the contributions 
from the soft mode and the massless quark gases, respectively.
The left and right panels show the DPR below and above $T_c$, respectively.
}
\label{fig:RATE-omega}
\end{figure*}

In this section, we shall show the numerical results of the DPR~(\ref{eq:DPR})
and electric conductivity~(\ref{eq:conductivity}) near the QCD CP
calculated with the photon self-energy obtained in the previous section.

\begin{figure}[t]
  \centering
  \includegraphics[keepaspectratio, scale=0.6]{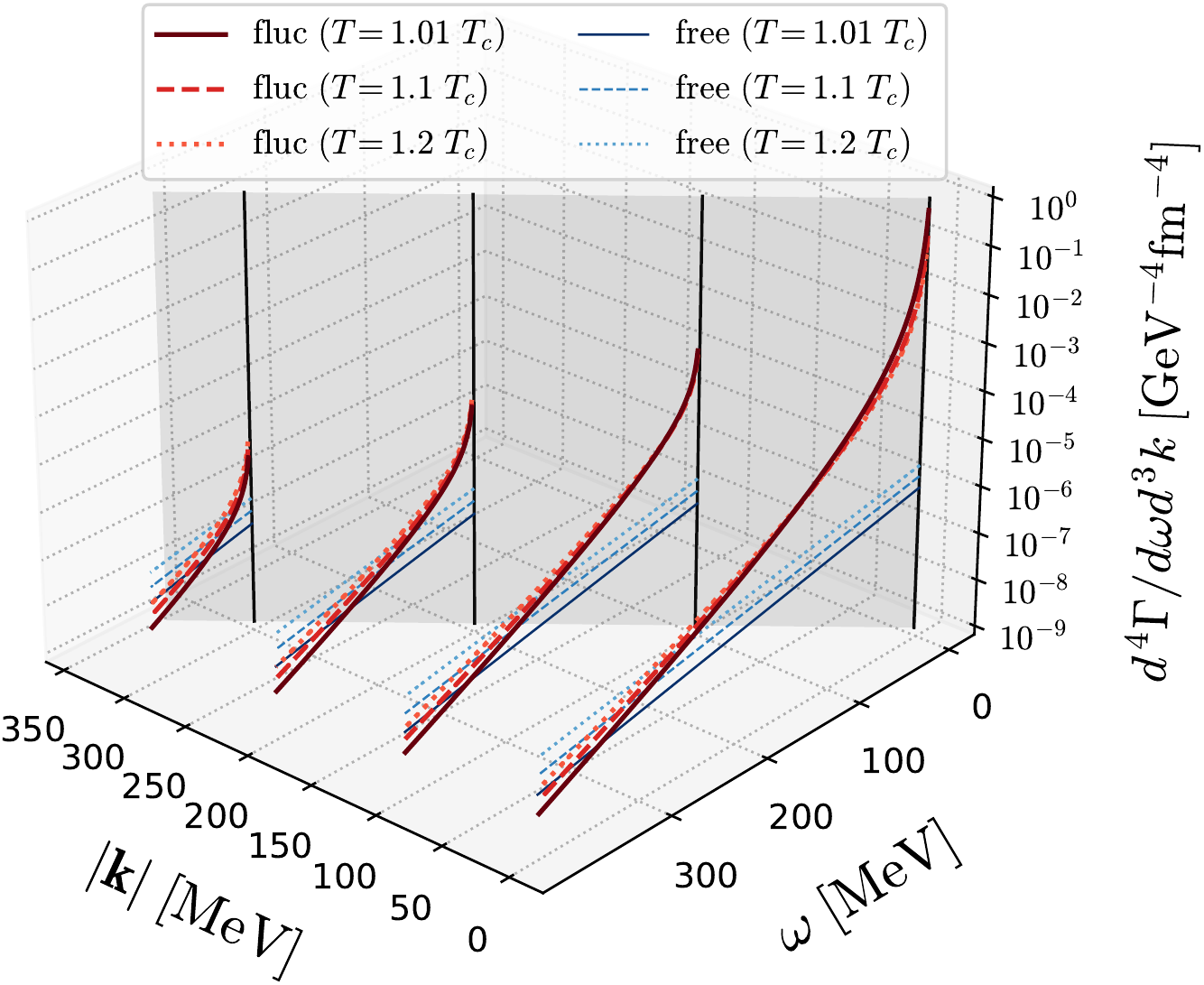}
  \caption{
DPR $d^4\Gamma / d\omega d^3k$ with the finite momentum $\bm{k}$
for $T/T_c = 1.01$, $1.1$ and $1.2$ at $\mu=\mu_c$. The gray surface shows the light-cone.}
  \label{fig:RATE-3d}
\end{figure}

We first show the DPR at $\bm{k}=\bm{0}$ 
at $\mu = \mu_c$ for several values of $T$ below (above) $T_c$
in the left (right) panel of Fig.~\ref{fig:RATE-omega}.
The red-thick lines show the contribution from $\tilde{\Pi}^{\mu\nu}_{\rm fluc} (k)$.
The total rate is given by the sum of the contributions from 
$\tilde{\Pi}^{\mu\nu}_{\rm fluc} (k)$ and $\tilde{\Pi}^{\mu\nu}_{\rm free} (k)$.
However, the latter is almost negligible at the QCD CP in the range of $\omega$ 
in the figure since ${\rm Im}\tilde{\Pi}^{\mu\nu}_{\rm free} (k)$ 
has a nonzero value only for $|\omega|>\sqrt{\bm{k}^2+4M^2}$, 
where $M\simeq185$~MeV at the QCD CP.
For a comparison, the DPR from the {\it massless} free-quark gas 
are shown by the blue-thin lines in the figure.
The figure shows that the DPR is enhanced significantly 
near the QCD CP by the soft modes
and well exceeds the case of the massless free-quark gas 
in the low energy region $\omega \lesssim 250~{\rm MeV}$.
The enhancement in the low energy region becomes 
more prominent as $T$ approaches $T_c$ from both sides of the temperature.
Taking a closer look at these results, one finds that 
the DPR increases monotonically in the left panel as $T$ approaches $T_c$,
while the $T$ dependence for $T>T_c$ shown in the right panel is not monotonous.
The latter can be accounted for by a competition of the effect of the soft modes 
and the kinematical temperature effect 
causing more thermal excitations of the soft modes at higher temperatures.
Figure \ref{fig:RATE-3d} shows the numerical results of the DPR 
at nonzero momentum for several $T$ above $T_c$.
One finds that the enhancement of the DPR 
is more prominent in the low-momentum region. 

\begin{figure*}[t]
    \begin{tabular}{cc}
      \begin{minipage}[t]{0.48\hsize}
        \centering
        \includegraphics[keepaspectratio, scale=0.4]{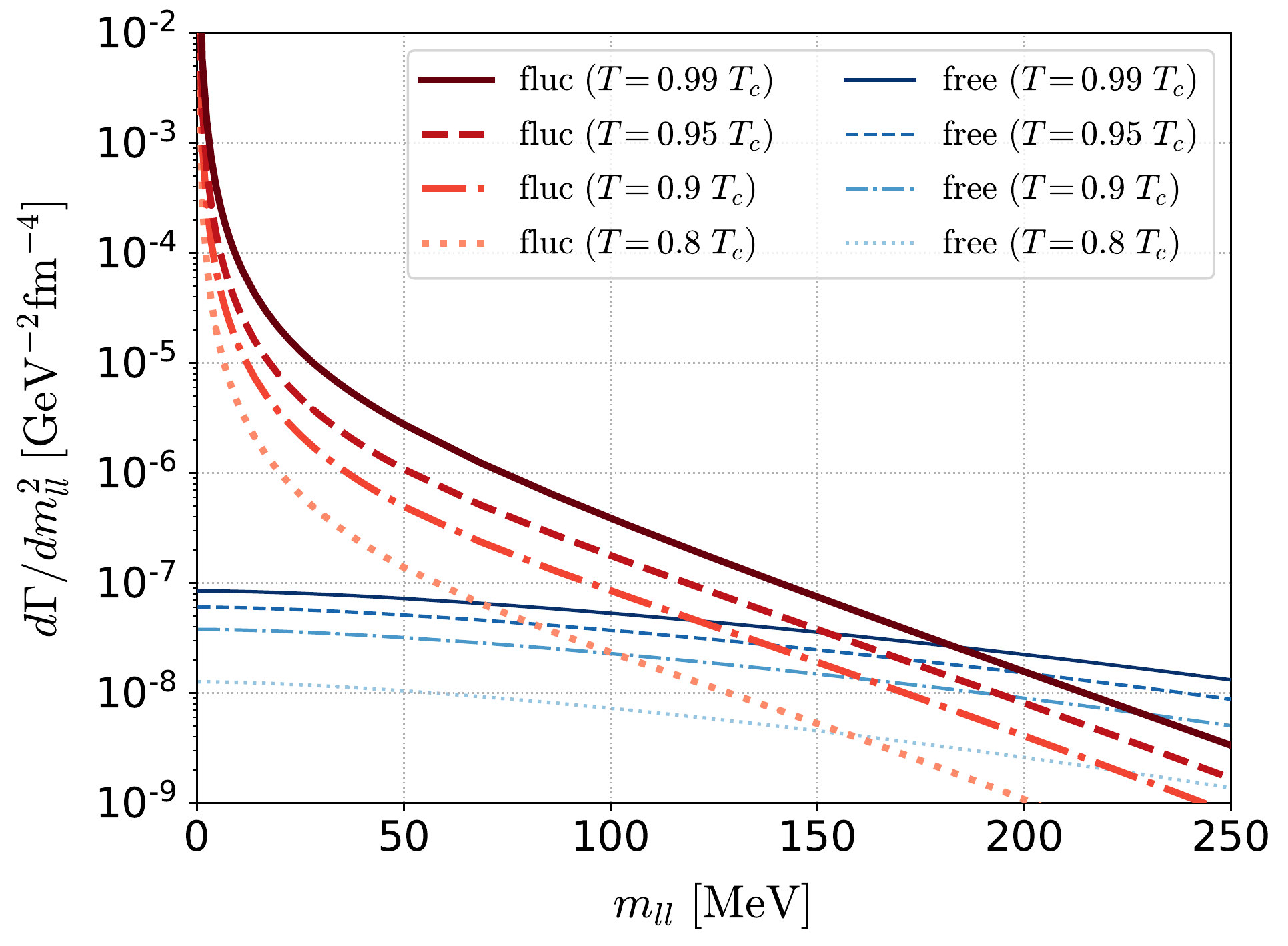}
      \end{minipage} 
      \begin{minipage}[t]{0.48\hsize}
        \centering
        \includegraphics[keepaspectratio, scale=0.4]{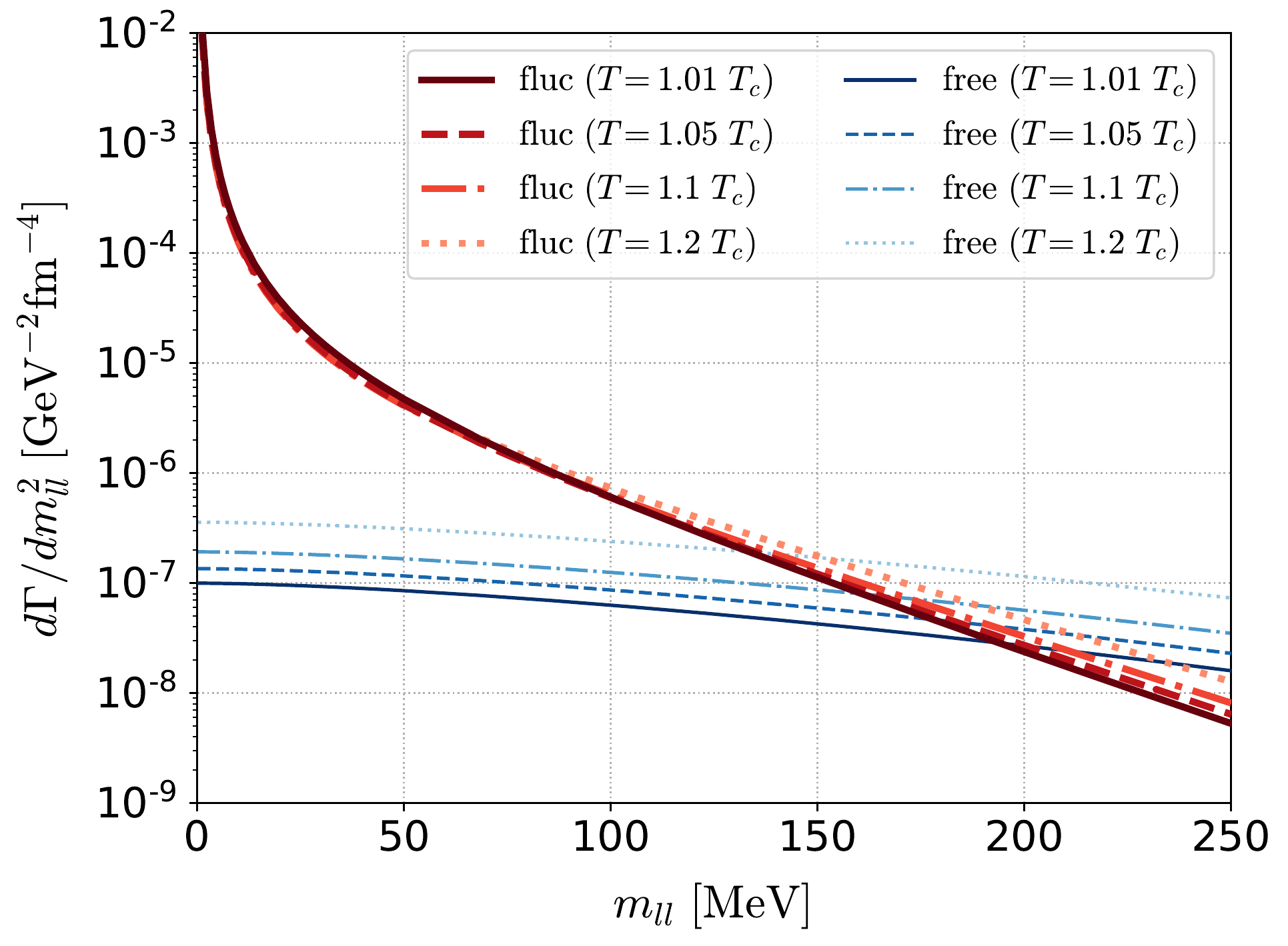}
      \end{minipage} 
     \end{tabular}
    \caption{Invariant-mass spectrum $d\Gamma / dm_{ll}^2$ at $\mu=\mu_c$.
{\bf Left}: For $T/T_c = 0.8$, $0.9$, $0.95$ and $0.99$.
{\bf Right}: For $T/T_c = 1.01$, $1.05$, $1.1$ and $1.20$.}
     \label{fig:RATE-M}
\end{figure*}

In the HIC experiments, the DPR is usually observed 
as a function of the invariant-mass $m_{ll}$,
\begin{align}
\frac{d\Gamma}{d m_{ll}^2} = \int d^3k \frac{1}{2\omega}
\frac{d^4 \Gamma}{d^4 k} \bigg|_{\omega=\sqrt{\bm{k}^2+m_{ll}^2}},
\label{eq:RATE-M}
\end{align}
to cancel out the effect of the flow.
In Fig.~\ref{fig:RATE-M}, we show the numerical results of Eq.~(\ref{eq:RATE-M})
for various values of $T$ at $\mu=\mu_c$. 
We find that the contribution of the soft modes is 
conspicuous in the low invariant-mass region $m_{ll} \lesssim 150~{\rm MeV}$.

\begin{figure*}[t]
     \begin{tabular}{cc}
      \begin{minipage}[t]{0.475\hsize}
        \centering
        \includegraphics[keepaspectratio, scale=0.42]{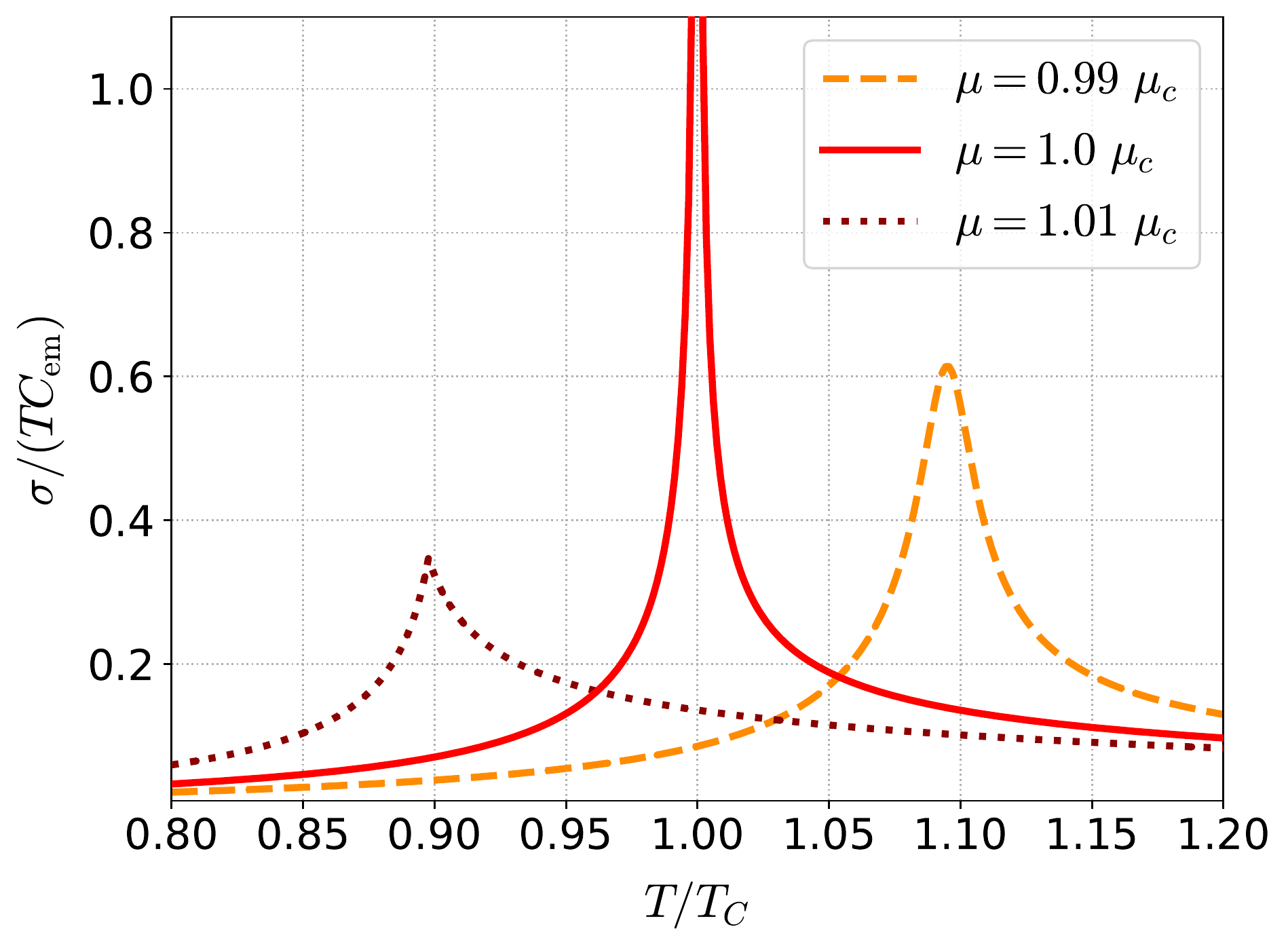}
      \end{minipage} 
      \begin{minipage}[t]{0.475\hsize}
        \centering
        \includegraphics[keepaspectratio, scale=0.42]{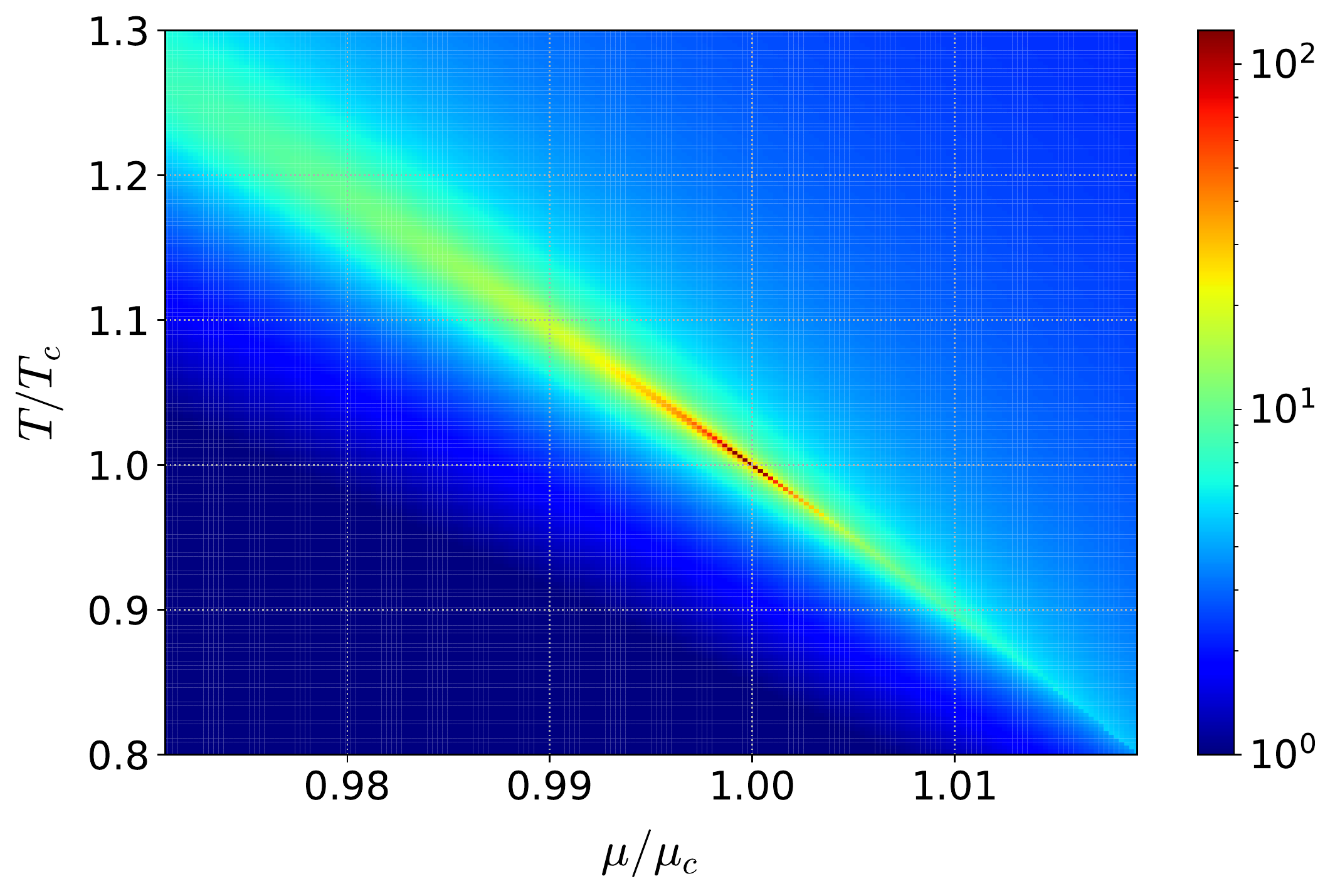}
      \end{minipage} 
     \end{tabular}
    \caption{
Electric conductivity $\sigma$ associated with the soft modes.
{\bf Left}: $T$ dependence of the $\sigma$.
The results of $\mu/\mu_c=0.99$, $1.0$ and $1.01$ are
the dashed, solid and dotted lines, repectively.
{\bf Right}: Contour plot of $\sigma$ in the $T$-$\mu$ plane around the QCD CP.}
    \label{fig:conductivity}
\end{figure*}

Finally, we show the behavior of the electric conductivity $\sigma$
near the QCD CP in Fig.~\ref{fig:conductivity}.
The left panel is the $T$ dependence of $\sigma$ at three values of $\mu$,
where $\sigma$ is normalized by $TC_{\rm em}$.
As expected from the infrared behavior of 
the soft modes in the critical region, the conductivity $\sigma$ 
tends to diverge near the CP.
In fact, it can be shown that $\sigma$ grows as $|T-T_c|^{-2/3}$
in the vicinity of the critical point in the present approximation,
as will be discussed in detail 
in the forthcoming publication~\cite{Nishimura_prep}.
At $\mu=0.99~\mu_c$, the conductivity is not divergent but only shows 
a prominent but finite peak at $T\simeq1.08~T_c$ in accordance 
with the crossover nature of the transition.
At $\mu=1.01~\mu_c$, the $\sigma$ shows a cusp-like behavior
reflecting the first-order nature of the phase transition at $T \simeq 0.9~T_c$.
The right panel of Fig.~\ref{fig:conductivity}
shows a contour plot of $\sigma/TC_{\rm em}$ on the $T$-$\mu$ plane.
One sees that the $\sigma$ has a significant excess along the critical lines of
the first-order phase and crossover transitions.

\section{Discussions}
Focusing on the collective soft modes 
the mass of which tends to vanish at the QCD CP, 
we have explored its effects on 
the dilepton production rate (DPR) and the electric conductivity near the QCD CP.
The contribution to these observables was
taken into account through the modification of the photon self-energy 
by the AL, MT and DOS terms, the inclusion of 
all of which is necessary to assure the WT identity.
We have shown that the DPR in the low energy and low invariant-mass regions 
is greatly enhanced due to the soft modes around the CP
in comparison with that of the massless free-quark gas.
We have also seen that the prominent enhancement of the
electric conductivity $\sigma$ occurs near the QCD CP due to the soft modes.
We plan to report on more detailed analyses 
on the possible anomalous transport properties 
including the electric conductivity and relaxation time 
near the QCD CP, as well as the phase boundary of the 2SC phase, elsewhere~\cite{Nishimura_prep}.

It is interesting to explore the phenomenological consequence of
the present findings in the HIC experiments.
If an anomalous enhancement of the DPR in the low mass region, 
say less than 150~MeV, should ever be detected, 
our result suggests that it may be the signal of the QCD CP.
The enhancement of the conductivity at the CP will also be observed 
in the HIC~\cite{Adamova:2019vkf}.
To identify the signal, however, it is important to disentangle 
the signal from other effects that induces a similar enhancement.
For example, in our previous study we have pointed out that a similar
enhancement of the DPR manifests itself near the phase boundary of the 2SC phase
due to the development of the diquark soft modes~\cite{Nishimura_prep}.
Other standard mechanisms due to medium effects, 
such as hadronic scenarios and the processes to be described 
by the perturbative QCD and so on
~\cite{Rapp:2009yu,Laine:2013vma,Ghiglieri:2014kma}, 
also bring about the enhancement at low mass region.
More detailed investigation of the invariant-mass spectrum 
of the DPR will be required to disentangle these effects.

Other important issues to be examined are the effects of dynamics. 
In the dynamical evolution of the HIC, the effect of the critical slowing down 
will modify the DPR around the QCD CP.
To deal with this effect, the analysis of the DPR in the real-time formalism 
with the time-dependent background medium is required.
It will also be important to understand the effects of the phase transitions 
on the bulk evolution of the medium~\cite{Savchuk:2022aev}.
For elucidating the production mechanisms and the respective characteristics, 
one would eventually need to recourse to some dynamical transport models~\cite{Bass:1998ca,Weil:2016zrk,Akamatsu:2018olk,Nara:2021fuu}.
These investigations are left for future study.
Nevertheless, we would like to emphasize that the production mechanism 
of the DPR through the soft modes is robust.

The measurements of the DPR in the low invariant-mass region 
$m_{ll} \lesssim 100-200~{\rm MeV}$ is also a challenge in the experimental side,
because di-electrons are contaminated by the Dalitz decay in this energy region.
In spite of these challenging demands, however,
it is encouraging that the future HIC programs 
in GSI and J-PARC-HI are designed to carry out 
high-statistical experiments~\cite{Galatyuk:2019lcf,Agarwal:2022ydl,Ozawa:2022sam}, 
and also that new technical developments 
are vigorously being made~\cite{Adamova:2019vkf}.

Finally, we remark that a complete description of the collective soft modes
around the QCD CP needs to incorporate 
the vector coupling~\cite{Kunihiro:1991qu,Kitazawa:2002jop} 
as well as the scalar couplings which was exclusively 
taken into account in the present study.
Such a more complete analyses constitutes one of the future tasks, 
which we hope to report somewhere in future.

\section*{Acknowledgements}
The authors thank Berndt Mueller, Hirotsugu Fujii and Akira Ohnishi 
for valuable comments.
T.~N. thanks JST SPRING (Grant No.~JPMJSP2138) and
Multidisciplinary PhD Program for Pioneering Quantum Beam Application.
This work was supported by JSPS KAKENHI 
(Grants No.~JP19K03872, No.~JP19H05598, No.~20H01903, No.~22K03619).

\bibliographystyle{elsarticle-num} 
\bibliography{reference.bib}

\begin{thebibliography}{10}
\expandafter\ifx\csname url\endcsname\relax
  \def\url#1{\texttt{#1}}\fi
\expandafter\ifx\csname urlprefix\endcsname\relax\def\urlprefix{URL }\fi
\expandafter\ifx\csname href\endcsname\relax
  \def\href#1#2{#2} \def\path#1{#1}\fi

\bibitem{Lovato:2022vgq}
A.~Lovato, et~al., {Long Range Plan: Dense matter theory for heavy-ion
  collisions and neutron stars} (11 2022).
\newblock \href {http://arxiv.org/abs/2211.02224} {\path{arXiv:2211.02224}}.

\bibitem{Stephanov:1998dy}
M.~A. Stephanov, K.~Rajagopal, E.~V. Shuryak, {Signatures of the tricritical
  point in QCD}, Phys. Rev. Lett. 81 (1998) 4816--4819.
\newblock \href {http://arxiv.org/abs/hep-ph/9806219}
  {\path{arXiv:hep-ph/9806219}}, \href
  {https://doi.org/10.1103/PhysRevLett.81.4816}
  {\path{doi:10.1103/PhysRevLett.81.4816}}.

\bibitem{Stephanov:1999zu}
M.~A. Stephanov, K.~Rajagopal, E.~V. Shuryak, {Event-by-event fluctuations in
  heavy ion collisions and the QCD critical point}, Phys. Rev. D 60 (1999)
  114028.
\newblock \href {http://arxiv.org/abs/hep-ph/9903292}
  {\path{arXiv:hep-ph/9903292}}, \href
  {https://doi.org/10.1103/PhysRevD.60.114028}
  {\path{doi:10.1103/PhysRevD.60.114028}}.

\bibitem{Hatta:2003wn}
Y.~Hatta, M.~A. Stephanov, {Proton number fluctuation as a signal of the QCD
  critical endpoint}, Phys. Rev. Lett. 91 (2003) 102003, [Erratum:
  Phys.Rev.Lett. 91, 129901 (2003)].
\newblock \href {http://arxiv.org/abs/hep-ph/0302002}
  {\path{arXiv:hep-ph/0302002}}, \href
  {https://doi.org/10.1103/PhysRevLett.91.102003}
  {\path{doi:10.1103/PhysRevLett.91.102003}}.

\bibitem{Minami:2009hn}
Y.~Minami, T.~Kunihiro, {Dynamical Density Fluctuations around QCD Critical
  Point Based on Dissipative Relativistic Fluid Dynamics -Possible fate of Mach
  cone at the critical point-}, Prog. Theor. Phys. 122 (2010) 881--910.
\newblock \href {http://arxiv.org/abs/0904.2270} {\path{arXiv:0904.2270}},
  \href {https://doi.org/10.1143/PTP.122.881} {\path{doi:10.1143/PTP.122.881}}.

\bibitem{Asakawa:2009aj}
M.~Asakawa, S.~Ejiri, M.~Kitazawa, {Third moments of conserved charges as
  probes of QCD phase structure}, Phys. Rev. Lett. 103 (2009) 262301.
\newblock \href {http://arxiv.org/abs/0904.2089} {\path{arXiv:0904.2089}},
  \href {https://doi.org/10.1103/PhysRevLett.103.262301}
  {\path{doi:10.1103/PhysRevLett.103.262301}}.

\bibitem{Asakawa:2015ybt}
M.~Asakawa, M.~Kitazawa, {Fluctuations of conserved charges in relativistic
  heavy ion collisions: An introduction}, Prog. Part. Nucl. Phys. 90 (2016)
  299--342.
\newblock \href {http://arxiv.org/abs/1512.05038} {\path{arXiv:1512.05038}},
  \href {https://doi.org/10.1016/j.ppnp.2016.04.002}
  {\path{doi:10.1016/j.ppnp.2016.04.002}}.

\bibitem{Hasanujjaman:2020zex}
M.~Hasanujjaman, G.~Sarwar, M.~Rahaman, A.~Bhattacharyya, J.-e. Alam,
  {Dynamical spectral structure of density fluctuation near the QCD critical
  point}, Eur. Phys. J. A 57~(10) (2021) 283.
\newblock \href {http://arxiv.org/abs/2008.03931} {\path{arXiv:2008.03931}},
  \href {https://doi.org/10.1140/epja/s10050-021-00589-3}
  {\path{doi:10.1140/epja/s10050-021-00589-3}}.

\bibitem{Galatyuk:2019lcf}
T.~Galatyuk, {Future facilities for high $\mu_B$ physics}, Nucl. Phys. A 982
  (2019) 163--169.
\newblock \href {https://doi.org/10.1016/j.nuclphysa.2018.11.025}
  {\path{doi:10.1016/j.nuclphysa.2018.11.025}}.

\bibitem{Agarwal:2022ydl}
K.~Agarwal, {Status of the Compressed Baryonic Matter (CBM) Experiment at
  FAIR}, Acta Phys. Polon. Supp. 16 (2023) 1--A142.
\newblock \href {http://arxiv.org/abs/2207.14585} {\path{arXiv:2207.14585}},
  \href {https://doi.org/10.5506/aphyspolbsupp.16.1-a142}
  {\path{doi:10.5506/aphyspolbsupp.16.1-a142}}.

\bibitem{Ozawa:2022sam}
K.~Ozawa, et~al., {The J-PARC heavy ion project}, EPJ Web Conf. 271 (2022)
  11004.
\newblock \href {https://doi.org/10.1051/epjconf/202227111004}
  {\path{doi:10.1051/epjconf/202227111004}}.

\bibitem{Fujii:2004jt}
H.~Fujii, M.~Ohtani, {Sigma and hydrodynamic modes along the critical line},
  Phys. Rev. D 70 (2004) 014016.
\newblock \href {http://arxiv.org/abs/hep-ph/0402263}
  {\path{arXiv:hep-ph/0402263}}, \href
  {https://doi.org/10.1103/PhysRevD.70.014016}
  {\path{doi:10.1103/PhysRevD.70.014016}}.

\bibitem{Son:2004iv}
D.~T. Son, M.~A. Stephanov, {Dynamic universality class of the QCD critical
  point}, Phys. Rev. D 70 (2004) 056001.
\newblock \href {http://arxiv.org/abs/hep-ph/0401052}
  {\path{arXiv:hep-ph/0401052}}, \href
  {https://doi.org/10.1103/PhysRevD.70.056001}
  {\path{doi:10.1103/PhysRevD.70.056001}}.

\bibitem{Yokota:2016tip}
T.~Yokota, T.~Kunihiro, K.~Morita, {Functional renormalization group analysis
  of the soft mode at the QCD critical point}, PTEP 2016~(7) (2016) 073D01.
\newblock \href {http://arxiv.org/abs/1603.02147} {\path{arXiv:1603.02147}},
  \href {https://doi.org/10.1093/ptep/ptw062} {\path{doi:10.1093/ptep/ptw062}}.

\bibitem{Yokota:2017uzu}
T.~Yokota, T.~Kunihiro, K.~Morita, {Tachyonic instability of the scalar mode
  prior to the QCD critical point based on the functional renormalization-group
  method in the two-flavor case}, Phys. Rev. D 96~(7) (2017) 074028.
\newblock \href {http://arxiv.org/abs/1707.05520} {\path{arXiv:1707.05520}},
  \href {https://doi.org/10.1103/PhysRevD.96.074028}
  {\path{doi:10.1103/PhysRevD.96.074028}}.

\bibitem{Nishimura:2022mku}
T.~Nishimura, M.~Kitazawa, T.~Kunihiro, {Anomalous enhancement of dilepton
  production as a precursor of color superconductivity}, PTEP 2022~(9) (2022)
  093D02.
\newblock \href {http://arxiv.org/abs/2201.01963} {\path{arXiv:2201.01963}},
  \href {https://doi.org/10.1093/ptep/ptac100}
  {\path{doi:10.1093/ptep/ptac100}}.

\bibitem{Voskresensky:2003wd}
D.~N. Voskresensky, {Fluctuations of the color superconducting order parameter
  in heated and dense quark matter} (6 2003).
\newblock \href {http://arxiv.org/abs/nucl-th/0306077}
  {\path{arXiv:nucl-th/0306077}}.

\bibitem{Kitazawa:2001ft}
M.~Kitazawa, T.~Koide, T.~Kunihiro, Y.~Nemoto, {Precursor of color
  superconductivity in hot quark matter}, Phys. Rev. D 65 (2002) 091504.
\newblock \href {http://arxiv.org/abs/nucl-th/0111022}
  {\path{arXiv:nucl-th/0111022}}, \href
  {https://doi.org/10.1103/PhysRevD.65.091504}
  {\path{doi:10.1103/PhysRevD.65.091504}}.

\bibitem{Kitazawa:2003cs}
M.~Kitazawa, T.~Koide, T.~Kunihiro, Y.~Nemoto, {Pseudogap of color
  superconductivity in heated quark matter}, Phys. Rev. D 70 (2004) 056003.
\newblock \href {http://arxiv.org/abs/hep-ph/0309026}
  {\path{arXiv:hep-ph/0309026}}, \href
  {https://doi.org/10.1103/PhysRevD.70.056003}
  {\path{doi:10.1103/PhysRevD.70.056003}}.

\bibitem{Kitazawa:2005vr}
M.~Kitazawa, T.~Koide, T.~Kunihiro, Y.~Nemoto, {Pre-critical phenomena of
  two-flavor color superconductivity in heated quark matter: Diquark-pair
  fluctuations and non-Fermi liquid behavior}, Prog. Theor. Phys. 114 (2005)
  117--155.
\newblock \href {http://arxiv.org/abs/hep-ph/0502035}
  {\path{arXiv:hep-ph/0502035}}, \href {https://doi.org/10.1143/PTP.114.117}
  {\path{doi:10.1143/PTP.114.117}}.

\bibitem{AL:1968}
L.~Aslamazov, A.~Larkin, Soviet solid state 10, 875 (1968), Phys. Lett. A 26
  (1968) 238.

\bibitem{Maki:1968}
K.~Maki, Critical fluctuation of the order parameter in a superconductor. {I},
  Progress of Theoretical Physics 40~(2) (1968) 193--200.

\bibitem{Thompson:1968}
R.~S. Thompson, Microwave, flux flow, and fluctuation resistance of dirty
  type-{II} superconductors, Physical Review B 1~(1) (1970) 327.

\bibitem{book_Larkin}
A.~Larkin, A.~Varlamov, Fluctuation phenomena in superconductors, Springer,
  2008.

\bibitem{Hatsuda:1994pi}
T.~Hatsuda, T.~Kunihiro, {QCD phenomenology based on a chiral effective
  Lagrangian}, Phys. Rept. 247 (1994) 221--367.
\newblock \href {http://arxiv.org/abs/hep-ph/9401310}
  {\path{arXiv:hep-ph/9401310}}, \href
  {https://doi.org/10.1016/0370-1573(94)90022-1}
  {\path{doi:10.1016/0370-1573(94)90022-1}}.

\bibitem{Fujii:2003bz}
H.~Fujii, {Scalar density fluctuation at critical end point in NJL model},
  Phys. Rev. D 67 (2003) 094018.
\newblock \href {http://arxiv.org/abs/hep-ph/0302167}
  {\path{arXiv:hep-ph/0302167}}, \href
  {https://doi.org/10.1103/PhysRevD.67.094018}
  {\path{doi:10.1103/PhysRevD.67.094018}}.

\bibitem{Thouless}
D.~J. Thouless, Perturbation theory in statistical mechanics and the theory of
  superconductivity, Annals of Physics 10~(4) (1960) 553--588.

\bibitem{book_Kapusta}
J.~I. Kapusta, C.~Gale, {Finite-temperature field theory: Principles and
  applications}, Cambridge Monographs on Mathematical Physics, Cambridge
  University Press, 2011.
\newblock \href {https://doi.org/10.1017/CBO9780511535130}
  {\path{doi:10.1017/CBO9780511535130}}.

\bibitem{Nishimura_prep}
T.~Nishimura, M.~Kitazawa, T.~Kunihiro, in preparation.

\bibitem{Adamova:2019vkf}
D.~Adamov\'a, et~al., {A next-generation LHC heavy-ion experiment} (1 2019).
\newblock \href {http://arxiv.org/abs/1902.01211} {\path{arXiv:1902.01211}}.

\bibitem{Rapp:2009yu}
R.~Rapp, J.~Wambach, H.~van Hees, {The Chiral Restoration Transition of QCD and
  Low Mass Dileptons}, Landolt-Bornstein 23 (2010) 134.
\newblock \href {http://arxiv.org/abs/0901.3289} {\path{arXiv:0901.3289}},
  \href {https://doi.org/10.1007/978-3-642-01539-7_6}
  {\path{doi:10.1007/978-3-642-01539-7_6}}.

\bibitem{Laine:2013vma}
M.~Laine, {NLO thermal dilepton rate at non-zero momentum}, JHEP 11 (2013) 120.
\newblock \href {http://arxiv.org/abs/1310.0164} {\path{arXiv:1310.0164}},
  \href {https://doi.org/10.1007/JHEP11(2013)120}
  {\path{doi:10.1007/JHEP11(2013)120}}.

\bibitem{Ghiglieri:2014kma}
J.~Ghiglieri, G.~D. Moore, {Low Mass Thermal Dilepton Production at NLO in a
  Weakly Coupled Quark-Gluon Plasma}, JHEP 12 (2014) 029.
\newblock \href {http://arxiv.org/abs/1410.4203} {\path{arXiv:1410.4203}},
  \href {https://doi.org/10.1007/JHEP12(2014)029}
  {\path{doi:10.1007/JHEP12(2014)029}}.

\bibitem{Savchuk:2022aev}
O.~Savchuk, A.~Motornenko, J.~Steinheimer, V.~Vovchenko, M.~Bleicher,
  M.~Gorenstein, T.~Galatyuk, {Enhanced dilepton emission from a phase
  transition in dense matter} (9 2022).
\newblock \href {http://arxiv.org/abs/2209.05267} {\path{arXiv:2209.05267}}.

\bibitem{Bass:1998ca}
S.~A. Bass, et~al., {Microscopic models for ultrarelativistic heavy ion
  collisions}, Prog. Part. Nucl. Phys. 41 (1998) 255--369.
\newblock \href {http://arxiv.org/abs/nucl-th/9803035}
  {\path{arXiv:nucl-th/9803035}}, \href
  {https://doi.org/10.1016/S0146-6410(98)00058-1}
  {\path{doi:10.1016/S0146-6410(98)00058-1}}.

\bibitem{Weil:2016zrk}
J.~Weil, et~al., {Particle production and equilibrium properties within a new
  hadron transport approach for heavy-ion collisions}, Phys. Rev. C 94~(5)
  (2016) 054905.
\newblock \href {http://arxiv.org/abs/1606.06642} {\path{arXiv:1606.06642}},
  \href {https://doi.org/10.1103/PhysRevC.94.054905}
  {\path{doi:10.1103/PhysRevC.94.054905}}.

\bibitem{Akamatsu:2018olk}
Y.~Akamatsu, M.~Asakawa, T.~Hirano, M.~Kitazawa, K.~Morita, K.~Murase, Y.~Nara,
  C.~Nonaka, A.~Ohnishi, {Dynamically integrated transport approach for
  heavy-ion collisions at high baryon density}, Phys. Rev. C 98~(2) (2018)
  024909.
\newblock \href {http://arxiv.org/abs/1805.09024} {\path{arXiv:1805.09024}},
  \href {https://doi.org/10.1103/PhysRevC.98.024909}
  {\path{doi:10.1103/PhysRevC.98.024909}}.

\bibitem{Nara:2021fuu}
Y.~Nara, A.~Ohnishi, {JAM mean-field update: mean-field effects on collective
  flow in high-energy heavy-ion collisions at $\sqrt{s_{NN}}=2-20$ GeV
  energies} (9 2021).
\newblock \href {http://arxiv.org/abs/2109.07594} {\path{arXiv:2109.07594}}.

\bibitem{Kunihiro:1991qu}
T.~Kunihiro, {Quark number susceptibility and fluctuations in the vector
  channel at high temperatures}, Phys. Lett. B 271 (1991) 395--402.
\newblock \href {https://doi.org/10.1016/0370-2693(91)90107-2}
  {\path{doi:10.1016/0370-2693(91)90107-2}}.

\bibitem{Kitazawa:2002jop}
M.~Kitazawa, T.~Koide, T.~Kunihiro, Y.~Nemoto, {Chiral and color
  superconducting phase transitions with vector interaction in a simple model},
  Prog. Theor. Phys. 108~(5) (2002) 929--951, [Erratum: Prog.Theor.Phys. 110,
  185--186 (2003)].
\newblock \href {http://arxiv.org/abs/hep-ph/0207255}
  {\path{arXiv:hep-ph/0207255}}, \href {https://doi.org/10.1143/PTP.108.929}
  {\path{doi:10.1143/PTP.108.929}}.

\end{thebibliography}

\end{document}